\documentclass[a4paper,fleqn,usenatbib]{mnras}

\usepackage{graphicx}	
\usepackage{amsmath}	
\usepackage{amssymb}	
\usepackage{url}
\usepackage{color,soul}

\title[A new model of the microwave polarized sky]{A new model of the microwave polarized sky for CMB experiments}

\author[Herv\'ias-Caimapo et al.]{
\parbox[t]{\textwidth}
{Carlos Herv\'ias-Caimapo$^{1}$\thanks{E-mail: carlos.herviascaimapo@postgrad.manchester.ac.uk}, Anna Bonaldi$^{1}$ and Michael L. Brown$^{1}$}
\vspace*{8pt} \\
$^{1}$Jodrell Bank Centre for Astrophysics, School of Physics \& Astronomy, University of Manchester, Oxford Road, Manchester M13 9PL, U.K.}

\date{Accepted XXX. Received YYY; in original form ZZZ}

\pubyear{2015}

\begin{document}
\label{firstpage}
\pagerange{\pageref{firstpage}--\pageref{lastpage}}
\maketitle

\begin{abstract}
We present a new model of the microwave sky in polarization that can be used to simulate data from CMB polarization experiments. We exploit the most recent results from the {\it Planck} satellite to provide an accurate description of the diffuse polarized foreground synchrotron and thermal dust emission. Our model can include the two mentioned foregrounds, and also a constructed template of Anomalous Microwave Emission (AME). Several options for the frequency dependence of the foregrounds can be easily selected, to reflect our uncertainties and to test the impact of different assumptions. Small angular scale features can be added to the foreground templates to simulate high-resolution observations.  

We present tests of the model outputs to show the excellent agreement with {\it Planck} and WMAP data. We determine the range within which the foreground spectral indices can be varied to be consistent with the current data. We also show forecasts for a high-sensitivity, high-resolution full-sky experiment such as the Cosmic ORigin Explorer (COrE). Our model is released as a python script that is quick and easy to use, available at \url{http://www.jb.man.ac.uk/~chervias}.
\end{abstract}

\begin{keywords}
methods: data analysis -- cosmic background radiation
\end{keywords}



\section{Introduction} 
Over the past decades, the temperature anisotropies of the Cosmic Microwave Background (CMB) have been an invaluable probe of the cosmological model \citep[e.g. ][]{2013ApJS20819H,2013PhRvD87j3012C,planck_2015_13}. The design of future CMB experiments is now driven by the goal of measuring accurately the polarization of the CMB, and searching for primordial polarization $B$-modes, a detection which would prove unequivocally the inflationary scenario. However, bright foreground emission due to our Galaxy can jeopardise this measurement, and accurate models of the polarization sky are needed \citep[see, e.g.][]{2009A&A...503..691B, 2012MNRAS.424.1914A, 2012PhRvD..85h3006E,2014MNRAS.444.1034B,2015PhRvL.114j1301B,2015arXiv150904714R}. 

Until recently, full-sky polarization maps of the Galactic emission were based on total intensity measurements and models of the polarization physical properties, angles and polarization fractions \citep[e.g. ][]{2011ASPC..449..187M,PSM2013,odea_2012}. However, the uncertainties in such modelling made it difficult to create polarization templates accurately reproducing the observed morphology in the sky. The recent release of the {\it Planck} data has improved this situation, by providing for the first time foreground maps extracted directly from the polarization data \citep{planck_2015_10}.  Before this information can be used to forecast future polarization experiments, however, it is necessary to overcome the limitations due to the {\it Planck} resolution and noise levels. Moreover, a suite of foreground models needs to be explored, to reflect the current uncertainties on polarized foregrounds. This is the goal of the current paper, where we deliver a new sky model of diffuse polarized emission in the microwave frequency range, based on the most up-to-date information from {\it Planck}.

In contrast to previous work \citep{PSM2013}, the model we present is not a comprehensive model that includes all point-like and diffuse emission in the microwave sky. Instead, we focus on diffuse polarized emission only and aim to provide a simpler and more flexible tool, to allow model selection for forecast purposes, as well as to test and debug data analysis methods on simulated data of varying complexity. We also introduce, for the first time, the capability to vary the foreground morphology for Monte-Carlo purposes. We believe that our model, which we provide as a python script, will be a useful tool for the CMB polarization community. 

The paper is organized as follows: In Sec. \ref{sec:model} we describe our sky model; in Sec. \ref{sec:observations} we describe the simulation procedure; in Sec. \ref{sec:comparison} we compare the outputs of our model with the most recent polarization data. In Sec. \ref{sec:forecast} we discuss the forecast and Monte-Carlo capabilities of our sky model and, finally, in Sec. \ref{sec:conclusions} we draw our conclusions.

\begin{figure}
    \centering
    \includegraphics[width=1\columnwidth]{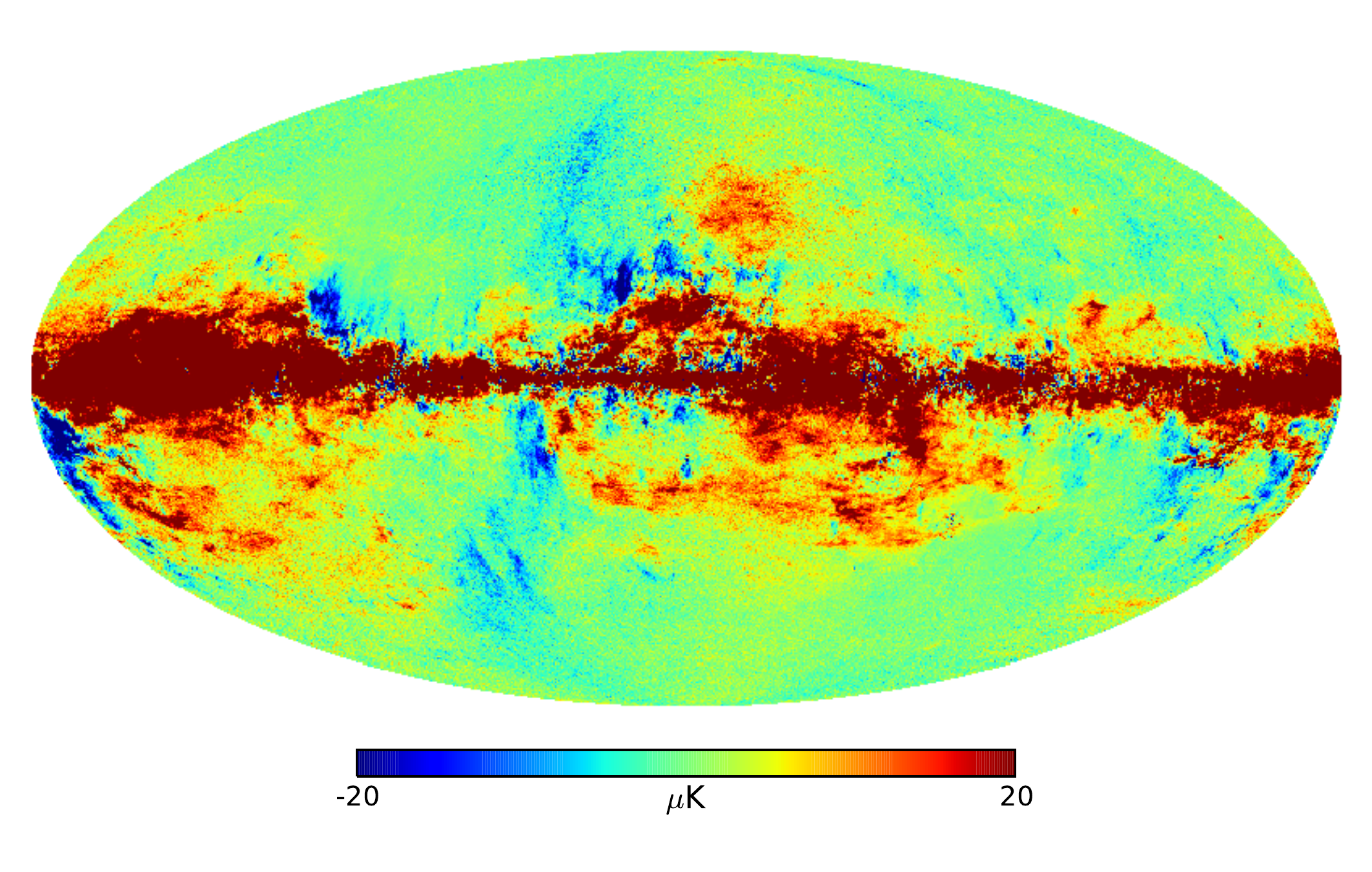}
    \includegraphics[width=1\columnwidth]{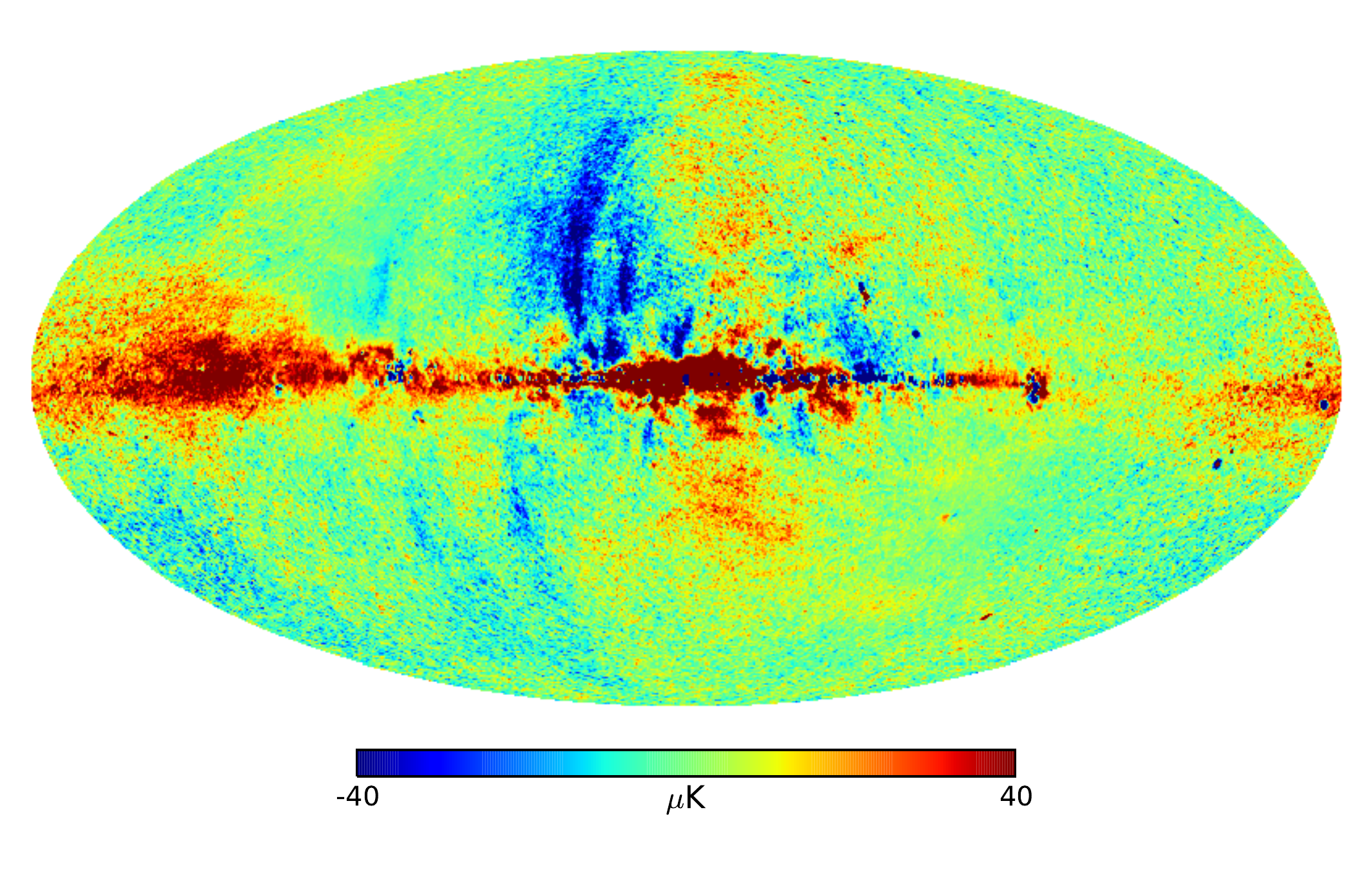}
    \caption{Top: template map of thermal dust $Q$ polarization intensity as derived in \citet{planck_2015_10} at 353\,GHz. Bottom: template map of synchrotron $Q$ polarization intensity as derived in \citet{planck_2015_10} at 30\,GHz. \label{fig:templates}}
\end{figure}

\section{Sky model}  \label{sec:model}
\subsection{CMB component}
The CMB is generated starting from a set of input $C_{\ell}$ power spectra from theory: $TT$ (the auto-spectrum of the total intensity), $EE$ and $BB$ (the two auto-spectra from the curl-free and divergence-free linear combination fields of polarization intensity), and $TE$ (the cross-spectrum between the total intensity and the polarization curl-free fields). These can be produced starting from a set of cosmological parameters, for example with the  \emph{CAMB} code \citep{howlett_2012}. The map is generated using the \emph{synfast} task of {\it HEALPix}\footnote{\url{http://healpix.jpl.nasa.gov}} \citep{gorski_2005}. It is then converted from thermodynamic to antenna temperature units at various frequencies with the usual black-body law with $T_{\rm CMB}=2.72548$ K.

\begin{figure}
  \centering
  \includegraphics[width=1\columnwidth]{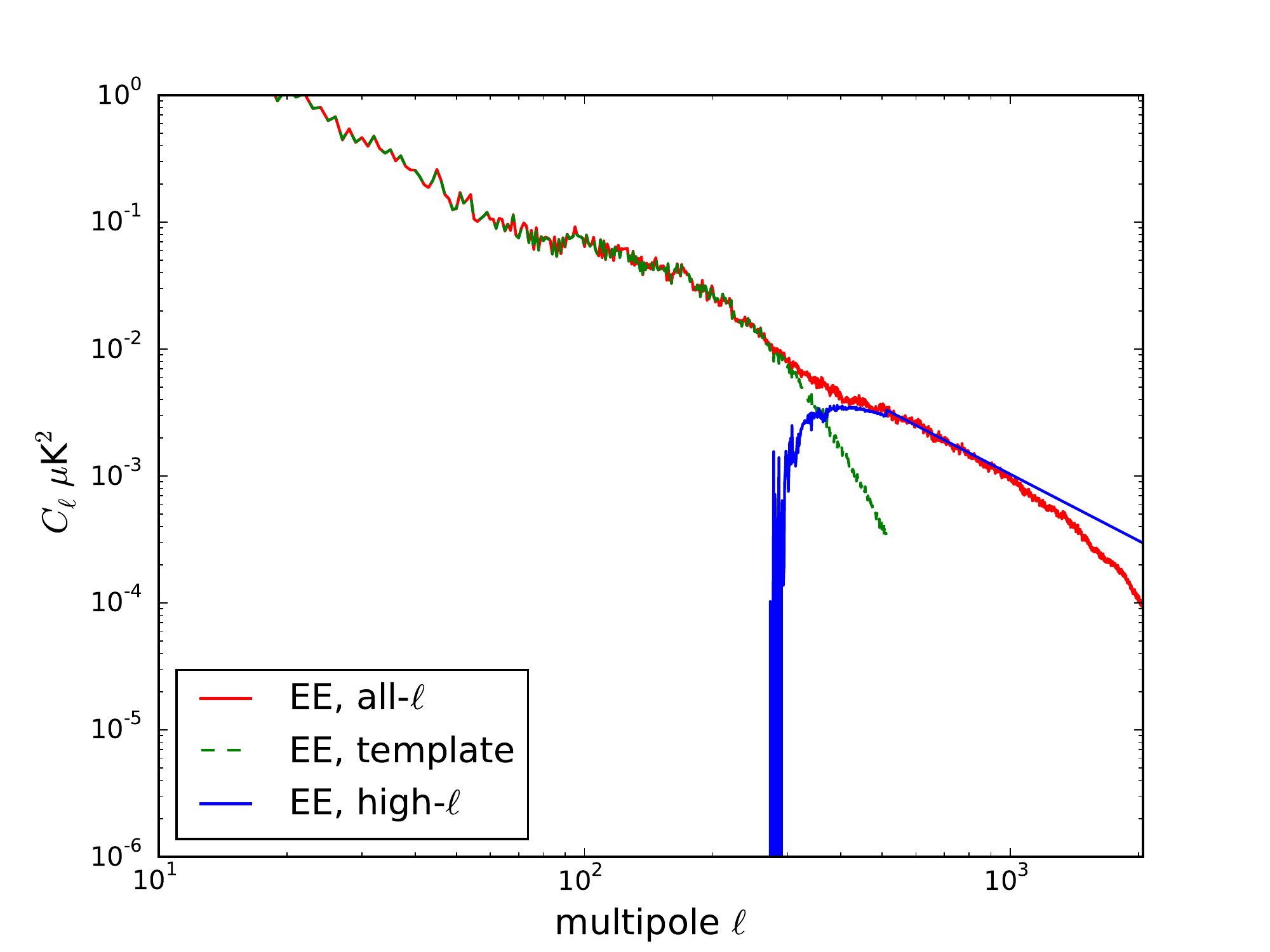}
  \includegraphics[width=1\columnwidth]{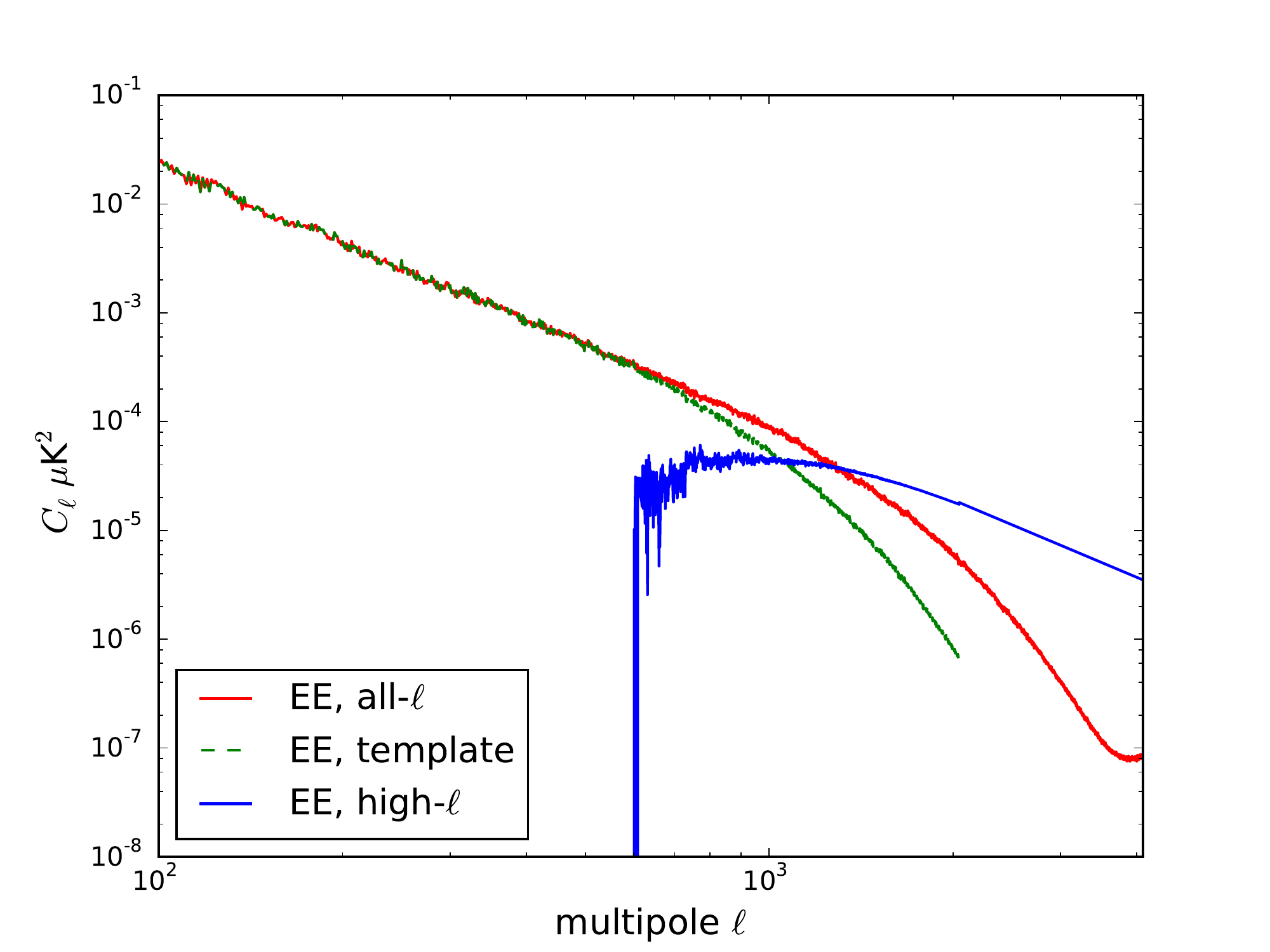}
  \caption{Synchrotron (top) and thermal dust (bottom) templates $EE$ power spectra. The $BB$ spectra is very similar, so it is omitted for clarity. The green curve corresponds to the original template. The blue curve is the high-$\ell$ extension, where the slope is extended to higher multipoles by a power-law fit. The red curve corresponds to the power spectra of the template map including the artificial high-$\ell$ features. Note that the red curve includes a $5'$ beam smoothing, whereas the blue curve is an extrapolation without smoothing. \label{fig:high-l-templates}}
\end{figure}

\subsection{Foreground templates} \label{sec:foregrounds_templates}
The simplest model of diffuse polarized foregrounds that is compatible with the observations has two Galactic polarized foregrounds: synchrotron and thermal dust.

We construct templates of these emission components based on the synchrotron and thermal dust polarization maps extracted from {\it Planck} observations with the Bayesian component separation method \emph{commander} \citep{planck_2015_10} and publicly available through the {\it Planck Legacy Archive}\footnote{\url{http://pla.esac.esa.int/pla/}}.

The synchrotron template has a reference frequency of 30\,GHz and a resolution of $40'$ Full Width at Half Maximum (FWHM). However, the pixel size of $\sim 14'$ (corresponding to $N_{\rm side}=256$) means that pixelization artefacts are visible on the maps. We eliminated these artefacts by resampling the map, upgrading it to $N_{\rm side}=512$ and smoothing it to a final $1^{\circ}$ resolution.

The dust template obtained by \cite{planck_2015_10} has a reference frequency of 353\,GHz and a resolution of $10'$ FWHM. Figure \ref{fig:templates} shows both $Q$ intensity templates.

\begin{figure}
    \centering
    \includegraphics[width=1\columnwidth]{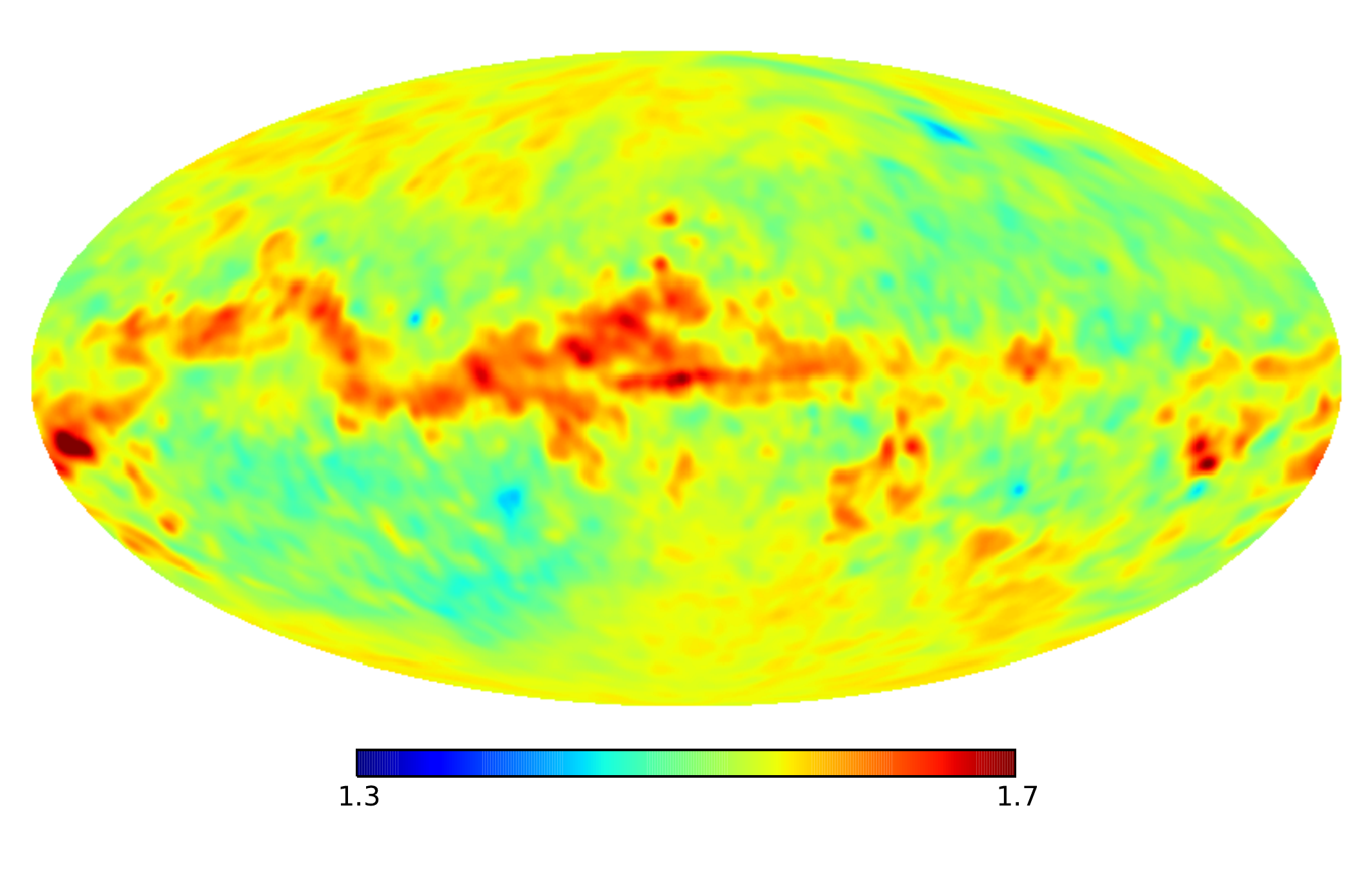}
    \includegraphics[width=1\columnwidth]{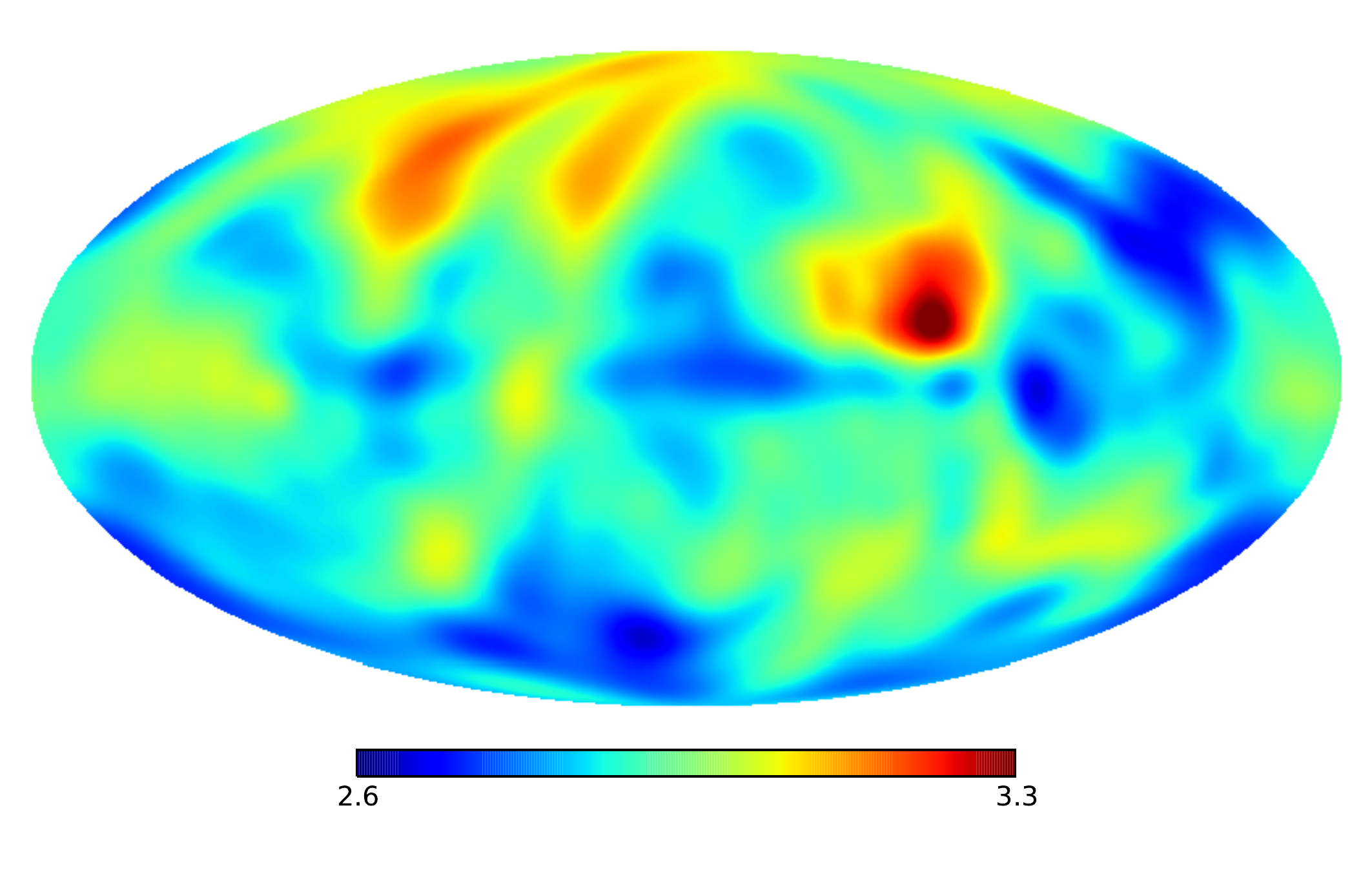}
    \caption{Top: map of thermal dust spectral indices based on \citet{planck_2015_10} and smoothed to $3^{\circ}$ FWHM to reduce noise. Bottom: map of synchrotron spectral indices based on \citet{giardino_2002} with $\beta_{\rm syn}$ increased to better fit the {\it Planck} frequency range (see text). \label{fig:spectral_indices}}
\end{figure}

\subsubsection{Adding high-$\ell$ features to the foreground templates} \label{sec:high-multipole} Since the synchrotron and thermal dust templates have finite resolution, they do not have power at small scales. In our models, we would like to simulate high-$\ell$ power since the real sky thermal dust and synchrotron emissions are expected to have these features. 

The approach that we follow is to generate a random map using a suitable power spectrum, based on the extrapolation of the power spectrum of the original map to higher multipoles.  Since this procedure involves a random realization, it can be also used to create variations over different foreground maps, for example for Monte Carlo purposes (this aspect is discussed in Sec. \ref{sec:forecast}). 

A common assumption is that the power spectrum of the foreground maps has a power-law behaviour in $\ell$. Figure \ref{fig:high-l-templates} (top) shows the $EE$ power spectra of the synchrotron template in green. The power-law behaviour is not a good approximation at the highest multipoles, where the slope flattens with respect to lower $\ell$s. We nonetheless adopted the power-law approximation and computed the best-fitting slope at the lowest multipoles.
 We used a least squares polynomial fit, that minimizes the difference between model and data, added in quadrature within a given multipole range. Our model is a straight line  in the $\log C_{\ell}$--$\log \ell$ space. This procedure also outputs the covariance matrix for the fit parameters, and we adopted the square root of the diagonal terms as errors on each of them. 
 
 For the $EE$ power spectrum, we fitted for the slope in the multipole interval $10 \leq \ell \leq 120$ and obtained a value of $-1.7\pm0.04$; for the $BB$ power spectrum, we fitted for it in the interval $4\leq\ell\leq40$ and obtained a flatter slope of $-1.4\pm0.05$. We obtained power spectra for the high-$\ell$ features as the difference between the original spectrum and its extrapolation computed using the best-fitting slope. This procedure creates a smooth high-$\ell$ power spectrum, shown in blue in Fig. \ref{fig:high-l-templates} (top). After this, we create a realization map with the artificial high-$\ell$ power spectrum, using \emph{synfast} and using a Gaussian beam appropriate for the resolution of the simulation. 

We finally multiply the resulting random map by a normalized version of the original template map. This reproduces the anisotropy of the foreground map (a Galactic plane mask in a sense), where regions in the Galactic plane are typically much brighter than at high latitudes. The high-$\ell$ random map is finally added to the original template, but multiplied by an amplitude chosen to give a continuous power spectrum at multipoles corresponding to the original beam. The power spectrum of the resulting map (for a $5'$ final resolution) is shown in Fig. \ref{fig:high-l-templates} (top) in red.

We follow the same procedure for the dust template. Fitting for the $EE$ and $BB$ slope in the range $60\leq\ell\leq600$ yields $-2.36\pm0.005$ and $-2.16\pm0.007$ respectively. Figure \ref{fig:high-l-templates} (bottom) shows the $EE$ power spectra for the dust template, for a final resolution of  $5'$. The colour code is the same as in Fig. \ref{fig:high-l-templates} (top). The slopes of the high-$\ell$ power spectrum, the beam and the amplitude of the high-$\ell$ maps are free parameters of the model and can be chosen to give different small-scale features, as needed for the simulation. 

\subsection{Baseline foreground model} \label{sec:baseline}
We model the frequency scaling of the dust and synchrotron components in antenna temperature as: 
\begin{eqnarray}
	T_{\rm A, dust}(\nu) &\propto& \nu^{\beta_{\rm dust}+1} [{\exp(h\nu/kT_{\rm d})-1}]^{-1} \label{eq:dust_law} \\
	T_{\rm A, syn}(\nu) &\propto& \nu^{-\beta_{\rm syn}} \text{,} \label{eq:syn_law}
\end{eqnarray}
where $h$ is the Planck constant, $k$ is the Boltzmann constant and $\nu$ is the frequency. The parameters $T_{\rm d}$, $\beta_{\rm dust}$ and $\beta_{\rm syn}$ are the dust temperature, dust spectral index and synchrotron spectral index, respectively.

The best-fitting values of \cite{planck_2015_10} are $T_{\rm d}=21$\,K and a spatially-varying dust spectral index with average value $\langle  \beta_{\rm dust}\rangle =1.53$ over the sky. For the synchrotron component \citet{planck_2015_10} uses a template spectrum obtained with the GALPROP code \citep{orlando_2013} instead of a power-law model; the slope of the spectrum between $\sim 19$ and $\sim 97$\,GHz corresponds to a  $\beta_{\rm syn} \sim 3.10$.  
For our baseline model we use spatially-constant parameters derived by the \cite{planck_2015_10} analysis: $T_{\rm d}=21$\,K, $\beta_{\rm dust} =1.53$ and $\beta_{\rm syn} = 3.10$. 

\subsection{Spatially-varying spectral indices} 
In order to add complexity to the models, we also considered using spatially varying spectral index maps for both dust and synchrotron emission. For thermal dust, we started from the map of best-fitting  spectral indices calculated using the temperature {\it Planck} maps from \emph{commander} in \citet{planck_2015_10}. This map has a resolution of $7.5'$ FWHM but it is very noisy. We therefore smoothed it to $3^{\circ}$. The final map of $\beta_{\rm dust}$ of our model is shown in the top panel of Fig.~\ref{fig:spectral_indices}. For our test model, we do not consider spatially varying $T_{\rm d}$, since there is a degeneracy between $\beta_{\rm dust}$ and $T_{\rm d}$. With no $\sim$THz data, it is very difficult to constrain both at the same time, so we only consider spatially varying $\beta_{\rm dust}$, which has more effect on the spectral law in the frequency range we consider. 

For synchrotron, we use the map of spectral indices by \citet{giardino_2002}. This map was derived using the full-sky map of synchrotron emission at 408\,MHz from \citet{haslam_1982}, the northern-hemisphere map at 1420\,MHz from \citet{reich_1986} and the southern-hemisphere map at 2326\,MHz from \citet{jonas_1998}. 
The \citet{giardino_2002} map has a resolution of $10^{\circ}$. 

One possible problem with the  \citet{giardino_2002} map is that is was derived at radio frequencies, where the synchrotron spectral index is typically flatter. We corrected for this effect by computing the expected steepening between $\sim 490$--$2120$\,MHz and $20$--30\,GHz using the same GALPROP template used in the {\it Planck} analysis and applying it to the \citet{giardino_2002} map. The result is shown in the bottom panel of Fig. \ref{fig:spectral_indices}; the steepening applied is $\Delta \beta_{\rm syn}=0.13$. The mean and standard deviation of this map are 2.9 and 0.1, respectively.

\subsection{Curved synchrotron spectral index and multiple thermal dust components} There is evidence that the synchrotron spectral law is not a constant power-law, instead having a curvature as the frequency increases \citep{kogut_2012}. In order to model this, we replace equation \ref{eq:syn_law} by
\begin{equation}
	T_{\rm A, syn}(\nu) \propto (\nu/\nu_0)^{-\beta_{\rm syn}+C\log(\nu/\nu_{\rm piv})} \text{,}
\end{equation}
where $C$ is the curvature amplitude, $\nu_0$ is the reference frequency of the synchrotron template and $\nu_{\rm piv}$ is a pivot frequency. Positive values of $C$ flatten, and negative ones steepen the spectral law for increasing frequency. For example, \cite{kogut_2007} finds a slight flattening of the spectrum with $C\sim0.3$ for $\nu_{\rm piv}=23$\,GHz for WMAP data.

The thermal dust spectral law might be better modelled using more than one modified black body, \citep[e.g.,][]{finkbeiner1999,meisner_2015}. The physical motivation is that different types of dust grains would be characterised by a different emission law. For this reason, we allow an arbitrary number of components, provided the user specifies $\beta_{\rm dust}$ (or a map of coordinate-dependent $\beta_{\rm dust}$), $T_{\rm d}$, and an amplitude $E_{\rm dust}$ for each component. We replace equation \ref{eq:dust_law} with
\begin{equation}
	T_{\rm A, dust}(\nu) \propto \sum_{\rm i=1}^{N_{\rm mbb}} E_{\rm dust,i}\, \nu^{\beta_{\rm dust,i}+1} [{\exp(h\nu/kT_{\rm d,i})-1}]^{-1} \text{,}
\end{equation}
where $N_{\rm mbb}$ is the number of modified black body components. We note that our parameterisation is equivalent to that in \cite{meisner_2015} once our $E_{{\rm dust}, i}$ is their $f_i q_i$. In that work, $q_i$ is a physical parameter describing the dust component, specifically the ratio of far-infrared emission cross-section to optical absorption cross-section. The parameter $f_i$ is the relative contribution (or fraction) of each component to the total (normalized such that $\sum_{i}^{N_{\rm mbb}} f_i = 1$). Our amplitude parameter $E_{{\rm dust}, i}$ accounts for both, and it is therefore a phenomenological, rather than a physical, parameter. For example, the best-fitting model (model 8) of \cite{finkbeiner1999} has two modified black body components that, in our parametrization, are described by $T_{\rm d,1}=9.4$\,K, $\beta_{\rm d,1}=1.67$, $T_{\rm d,2}=16.2$\,K, $\beta_{\rm d,2}=2.70$, and intensity ratios $E_{\rm dust,1}/E_{\rm dust,2}=0.49$.

\begin{figure}
	\centering
	\includegraphics[width=1\columnwidth]{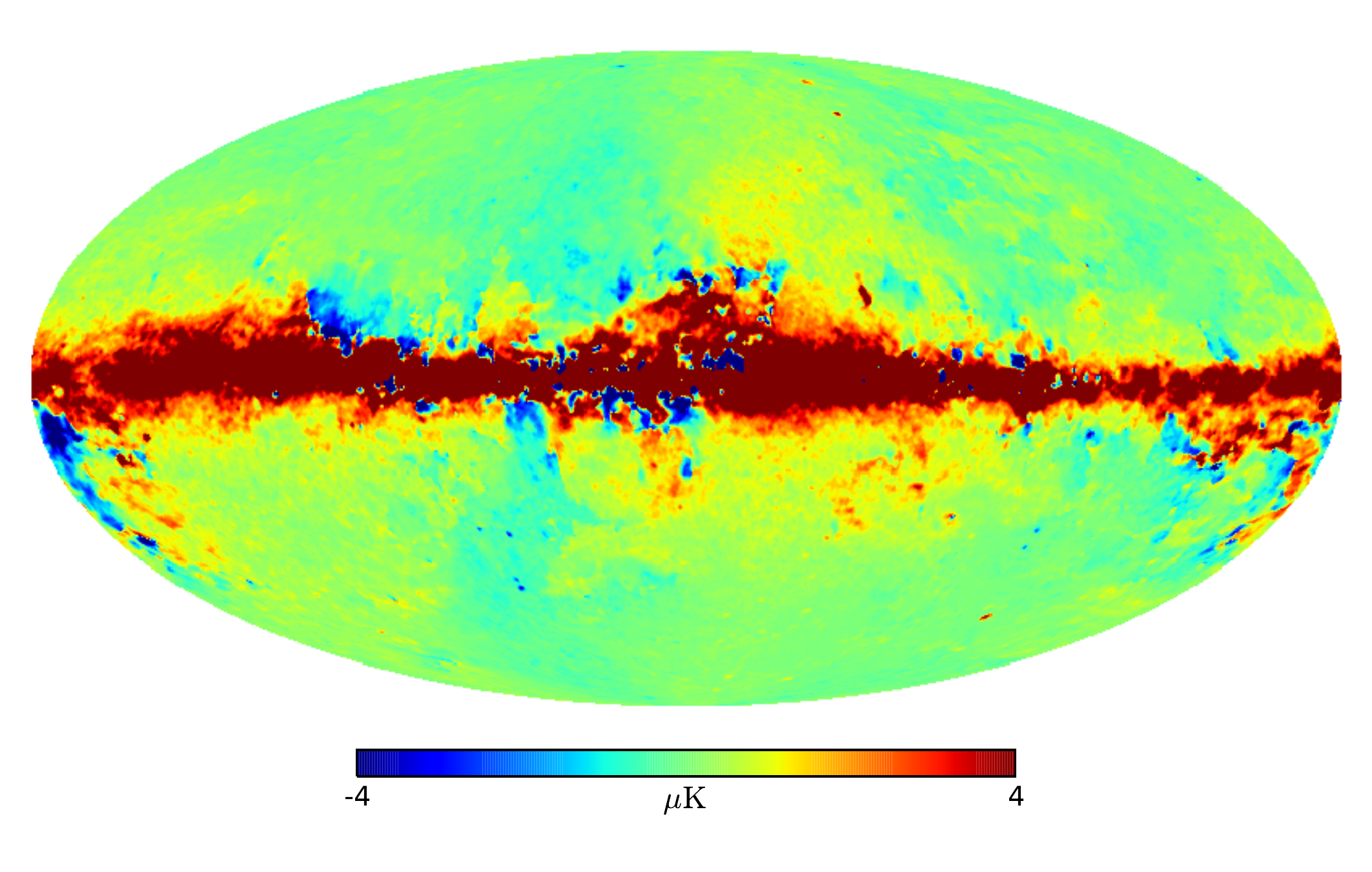}
	\caption{Template map of $Q$ AME component, as derived from the total intensity AME and the thermal dust polarization maps from \citet{planck_2015_10}. This map has $1^{\circ}$ resolution, a reference frequency of 23\,GHz and assumed a polarization fraction of 0.01. \label{fig:ame-template}}
\end{figure}

\begin{figure*}
	\centering
	\includegraphics[width=1.0\textwidth]{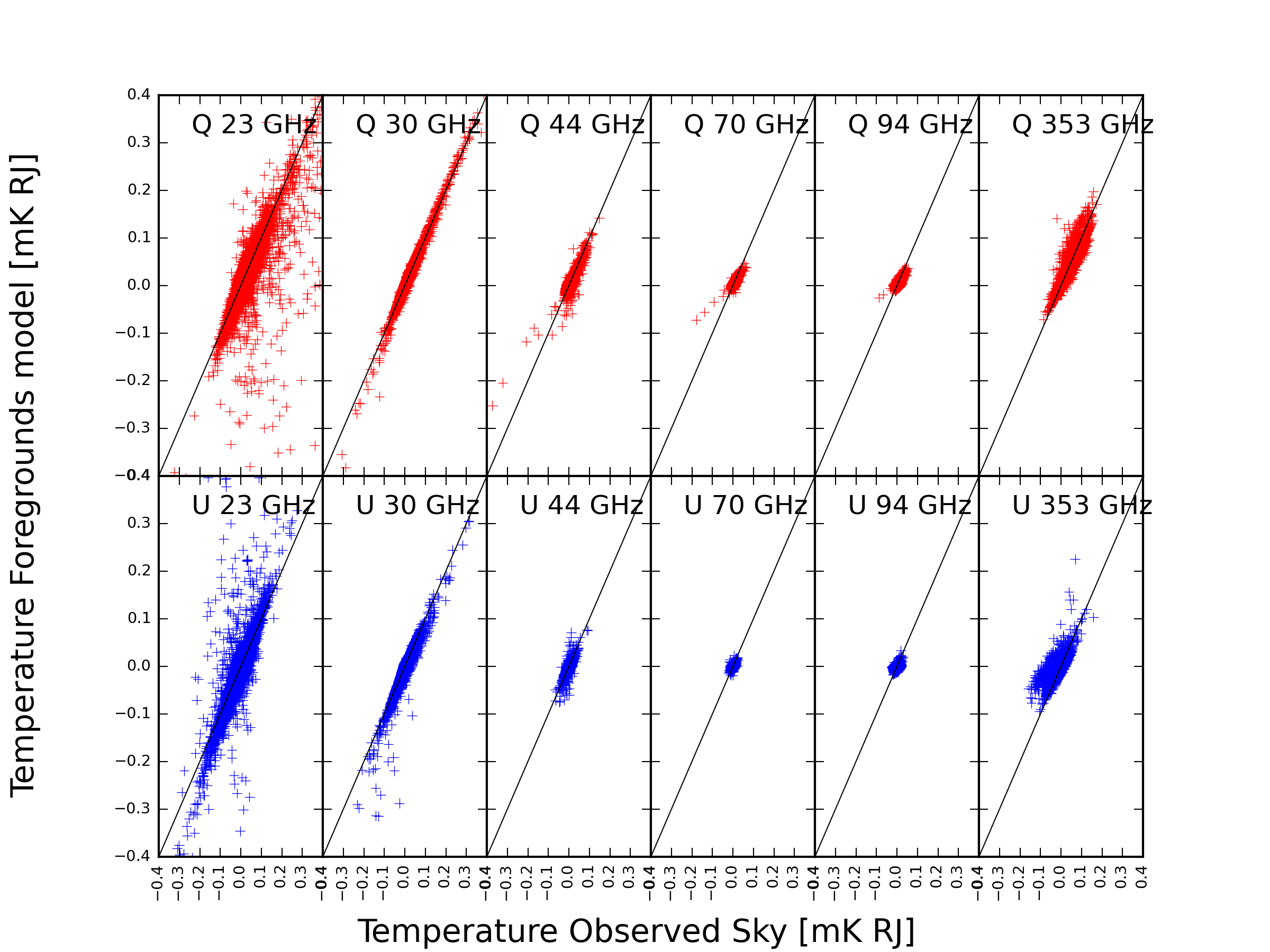}
	\caption{Scatter plot inside the Galactic plane $|b| \leq 20^{\circ}$. The $N_{\rm side}$ is 64. The top row corresponds to $Q$ intensity, and the bottom row to $U$ intensity. The black line represents the perfect one-to-one match. Both the templates and the observed sky were smoothed to a common $1^{\circ}$ resolution. \label{fig:scatter_plot} }
\end{figure*}

\subsection{Additional polarized components: Anomalous Microwave Emission (AME)} There is evidence that AME due to spinning dust is polarized, with a polarization fraction of few \% \citep{dickinson_2011,2015MNRAS.452.4169G}. We consider the polarization intensity of the AME as an additional feature to simulate observations by future experiments with a better accuracy. 

To construct our AME template we used the {\it Planck} 2015 total intensity AME template (with a resolution of $1^{\circ}$) and the thermal dust polarization maps. By assuming that the polarization angles for AME are the same as for the thermal dust, we can obtain polarization $Q$ and $U$ maps for AME as
\begin{eqnarray}
	Q_{\rm AME} &=& f_{\rm p, AME}\, T_{\rm AME}\, \cos(2\chi_{\rm TD}) \\
	U_{\rm AME} &=& f_{\rm p, AME}\, T_{\rm AME}\, \sin(2\chi_{\rm TD}) \text{,}
\end{eqnarray}
where $f_{\rm p,AME}$ is a spatially-constant polarization fraction (we used a default value of 1\%), $T_{\rm AME}$ is the total intensity AME template, and $\chi_{\rm TD}$ is the thermal dust polarization angle. The Pearson correlation coefficient between the thermal dust and the AME template (at $1^{\circ}$ resolution and $N_{\rm side}=64$) is $0.71 \pm 0.01$ ($Q$ map) and $0.73 \pm 0.01$ ($U$ map). The errors were calculated with jackknife resampling. Figure \ref{fig:ame-template} shows the $Q$ intensity of the constructed AME template, with a reference frequency of 23\,GHz.

As a spectral law, we adopt a parabola in the logarithmic flux-frequency space, proposed by \cite{bonaldi_2007}, given by
\begin{multline}
		\log(T_{\rm A,\nu}) = \text{const.} - \left[ \frac{m_{60}\log(\nu_{\rm max})}{\log(\nu_{\rm max}/60\text{GHz})} + 2 \right]\log(\nu) + \\
		\frac{m_{60}}{2\log(\nu_{\rm max}/60\text{GHz})}(\log(\nu))^2 \text{,}
\end{multline}
where the free parameters are $m_{60}$ (the slope at 60\,GHz) in the $\log(\nu)$-$\log(S)$ space, and $\nu_{\rm max}$ is the peak frequency (for the spectrum in flux units). We adopt as default values $\nu_{\rm max}=19$\,GHz, from \cite{planck_2015_10}, and 4.0 for $m_{60}$ from \cite{bonaldi_2007}.

\subsubsection{Adding dust-correlated high-$\ell$ features to the AME maps}
Similarly to what done for the synchrotron and thermal dust components in Sec.~\ref{sec:high-multipole}, the AME polarization maps can be upgraded in resolution by adding high-$\ell$ features. However, in this case we want the high-$\ell$ thermal dust and AME maps to exhibit the same level of correlation measured at low resolution. We therefore developed a special procedure for this case, that generates both dust and AME high-$\ell$ correlated random maps at the same time. 

We followed the procedure described in \cite{brown_2011}, which uses as input the spectra and cross-spectra of the set of correlated maps (in our case, dust $E$ and $B$, and AME $E$ and $B$). This information is used to generate 4 correlated $a_{\ell m}$ fields, which are finally transformed to $Q$ and $U$ with the {\it HEALPix alm2map} function. Extending what is described in Sec.~\ref{sec:high-multipole}, the spectra and cross-spectra for the high-$\ell$ maps are constructed by extrapolating those of the dust and AME templates to higher multipoles. The last step of our procedure, the modulation of the random high-$\ell$ maps with a mask enhancing the Galactic plane, is unchanged.

\begin{figure*}
    \centering
    \includegraphics[width=1\columnwidth]{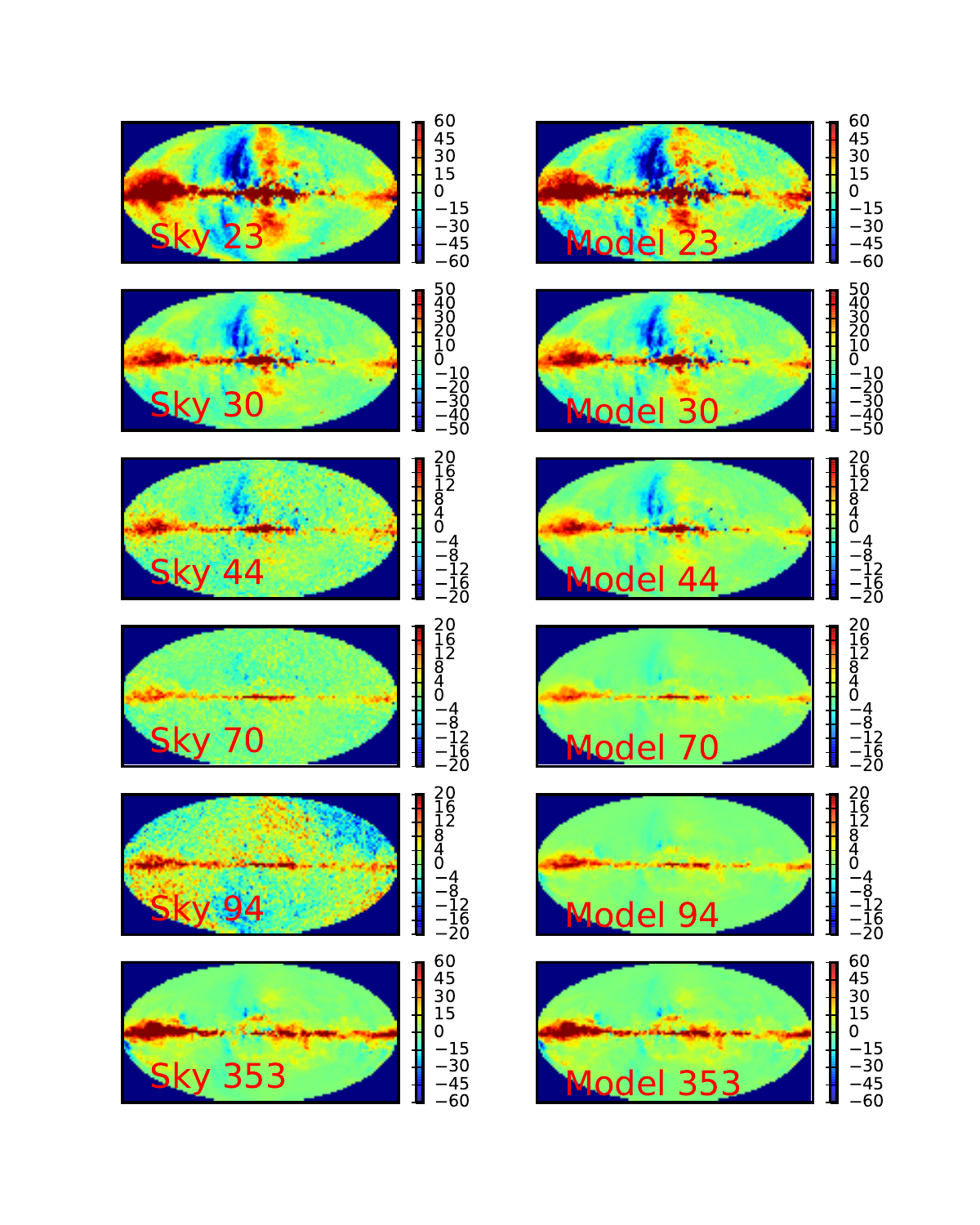}
    \includegraphics[width=1\columnwidth]{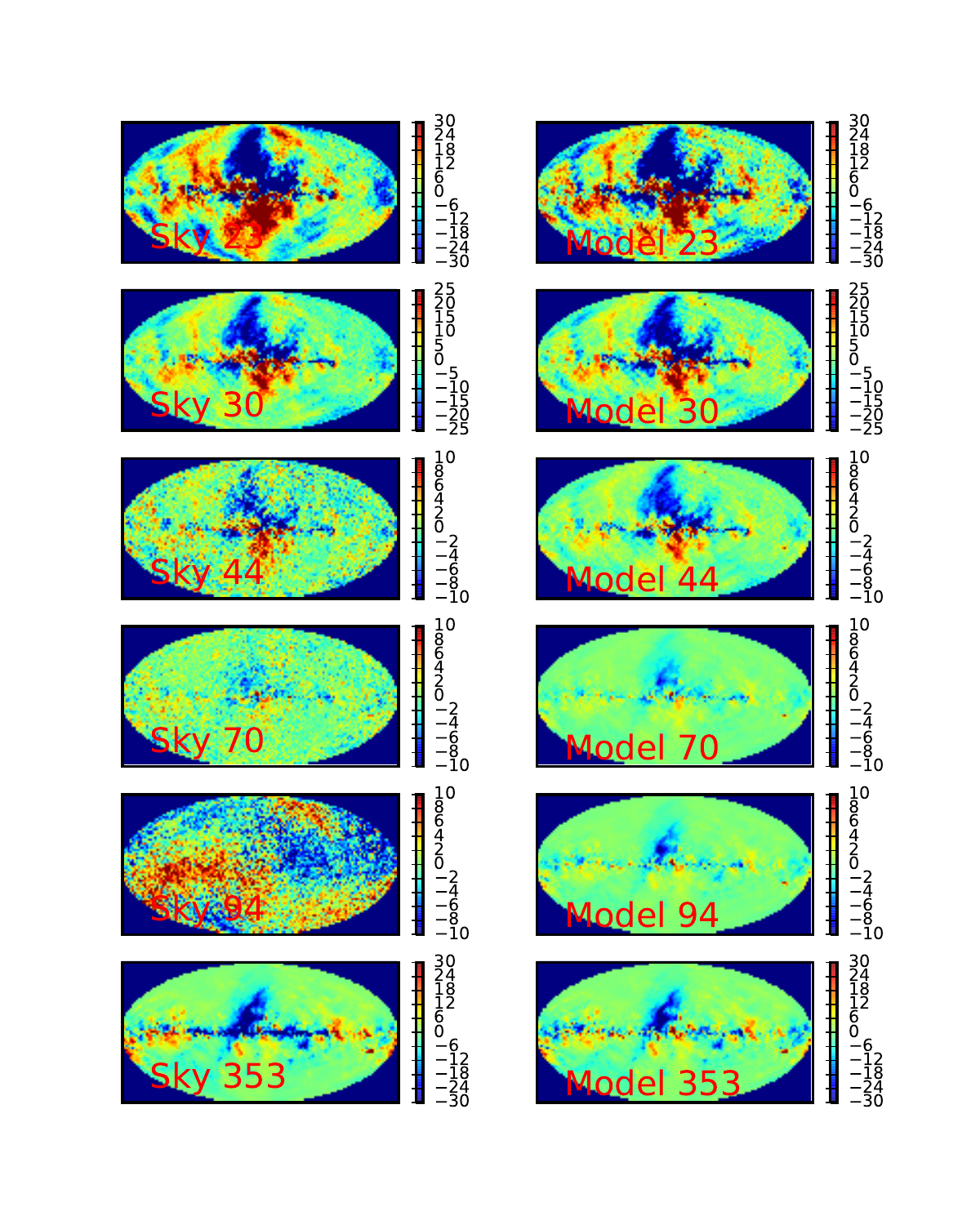}
    \caption{Maps comparison between the observed sky and the foreground model. Left: maps for $Q$ intensity. Right: maps for $U$ intensity. The rows are six bands, the left column corresponds to the observed sky and the right one to our foregrounds model. The units are $\mu{\rm K}_{\rm A}$. The maps are smoothed to a common $1^{\circ}$ resolution and degraded to $N_{\rm side}=32$ to suppress the noise for display purposes. \label{fig:maps_comparison}}
\end{figure*}

\section{Simulated observations of CMB polarization experiments} \label{sec:observations}

\subsection{Simulating the instrumental response} \label{sec:instrumenstal_response}
To simulate the observation of the microwave sky in polarization by a given experiment, we need to know the frequency bands of observation and, for each of the bands, the point-spread function and the noise level. Each of these properties can be simulated with different level of complexity, specified by the user, as detailed in the following.

The frequency response can be either simulated as a delta function (monochromatic response) or a more general transmission. In the more general case, the intensity of the sky component $i$ at the frequency band $\nu_j$ is given by:
\begin{equation}
	T_i(\nu_j) = \frac{\sum_{k} W_j(\nu_k) [Q/U]_{\rm ref} S_i(\nu_k)}{\sum_k W_j(\nu_k)} \text{,}
\end{equation}
where $S_i(\nu)$ is the spectral law of the component, $[Q/U]_{\rm ref}$ is the $Q$ or $U$ amplitude of the corresponding template, and $W_j(\nu_k)$ is the transmission of band $j$ for a set of frequencies $\nu_k$. In practice, when simulating a band response, the signal needs to be simulated for a set of frequencies $\nu_k$ and averaged over the entire band, with weights given by the transmission $W_j(\nu_k)$.

The effect of the instrumental resolution is simulated by convolving the maps with a Gaussian beam of specified FWHM. 

The noise can be modelled either as a uniform white noise, described by a unique rms value over all the sky, or as an anisotropic white noise specifying a map of rms varying in the sky. For {\it Planck}, we model this using the $3 \times 3$ noise covariance per pixel containing the Stokes parameter covariance elements $TT$, $QQ$, $UU$, $TQ$, $TU$, and $QU$. In this case, for each pixel, a Cholesky decomposition is performed over the covariance matrix; the diagonal elements of the decomposition finally yield the standard deviations per pixel for $T$, $Q$, and $U$, respectively.

\subsection{Simulation procedure} Once the experiment is specified, by means of a set of frequencies, resolution and noise, the simulation procedure is the following:
\begin{itemize}
\item A CMB map is generated using {\it synfast}, up to a resolution equal to $\theta_{*}$, which should be at least equal to the smallest instrumental beam of the experiment.
\item High-$\ell$ features are optionally added to the synchrotron, dust and/or AME templates up to a resolution $\theta_{*}$ (if $\theta_{*}$ is smaller than the intrinsic resolution of the template).
\item The CMB map and foreground templates are scaled in intensity according to the frequency behaviour to each frequency band, added together and smoothed to match the resolution appropriate for that channel.
\item A noise map is generated and added to the frequency band for each channel to obtain the simulated frequency map.
\end{itemize}

The outputs are the frequency maps, but also the component maps at all required frequencies. Some of the components can be easily deactivated to obtain noise-only, signal-only, foreground-only or CMB-only simulations, for example for Monte Carlo purposes. 

\begin{figure}
	\centering
	\includegraphics[width=1\columnwidth]{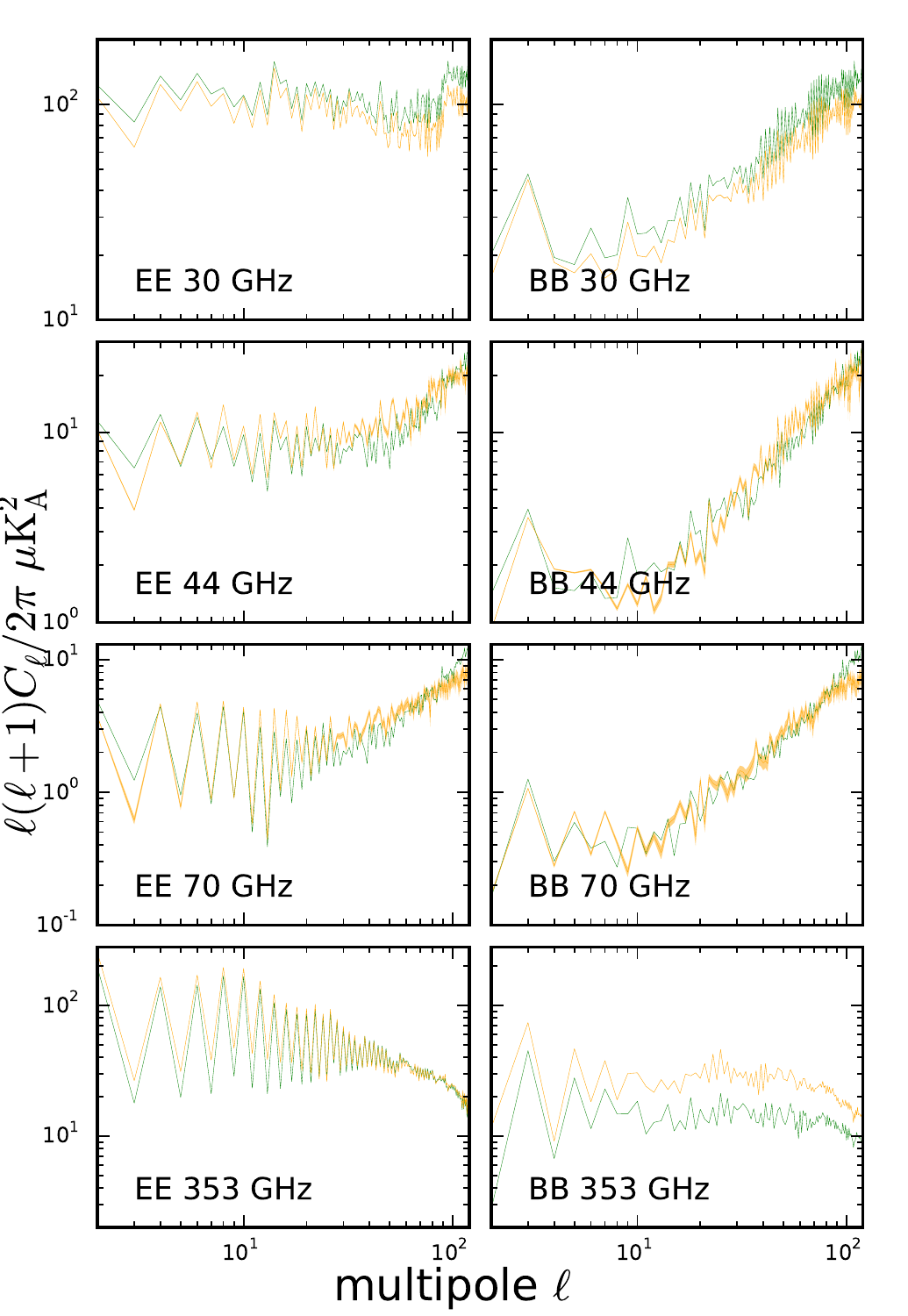}
	\caption{$EE$ (left) and $BB$ (right) full-sky power spectra comparison between the complete model (foregrounds+noise+CMB, in green) and the observed sky (in orange) in four {\it Planck} bands, for $N_{\rm side}=256$. The error for the observations is plotted as the orange shaded region. The full-sky maps are smoothed to a common $1^{\circ}$ resolution. \label{fig:power_spectra_complete_model_planck}}
\end{figure}
\begin{figure}
	\centering
	\includegraphics[width=1\columnwidth]{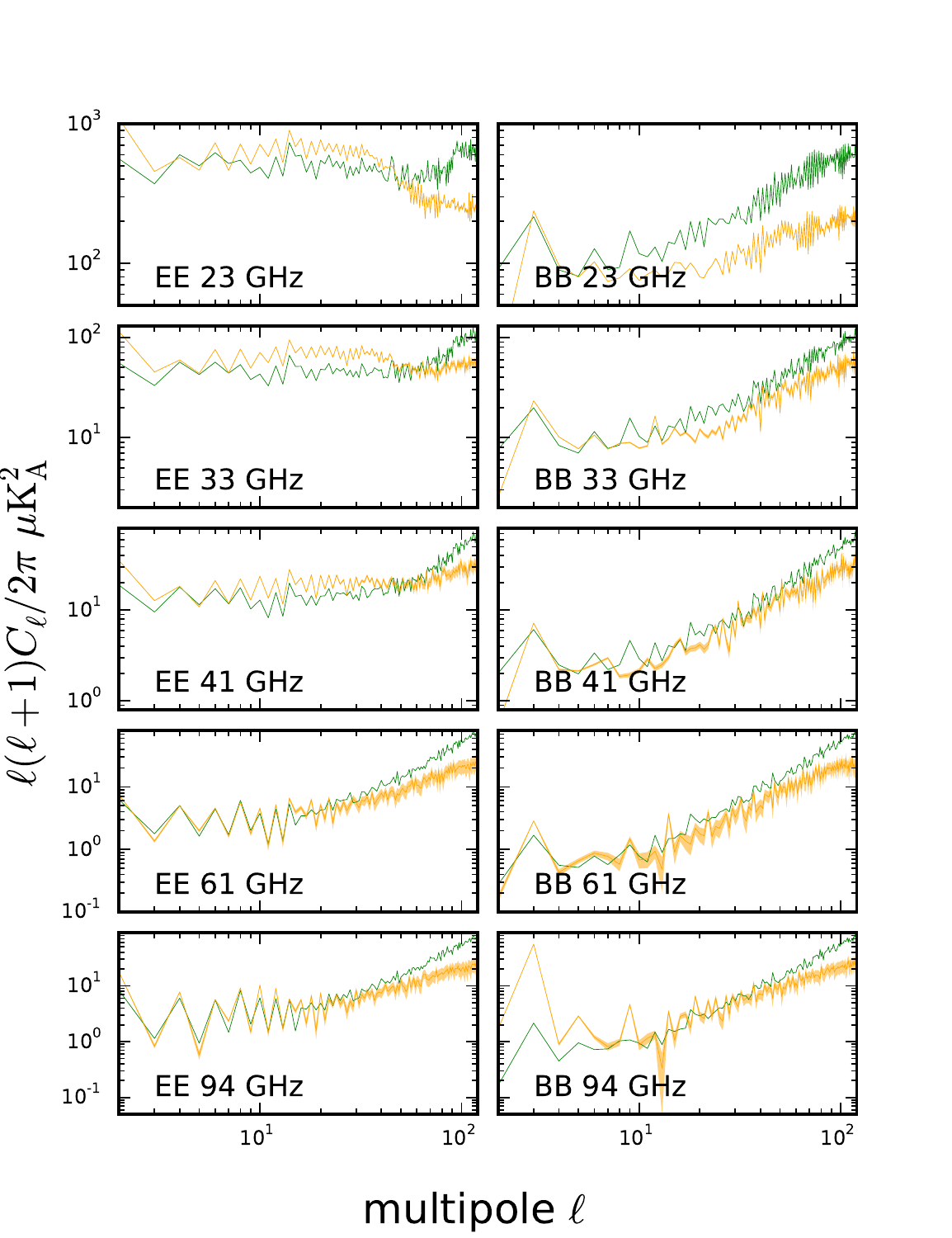}
	\caption{$EE$ (left) and $BB$ (right) full-sky power spectra comparison for the five WMAP bands. The error for the observations is plotted as the orange shaded region. The full-sky maps are smoothed to a common $1^{\circ}$ resolution. The convention is the same as in Fig.~\ref{fig:power_spectra_complete_model_planck}. \label{fig:power_spectra_complete_model_wmap}}
\end{figure} 

\section{Comparison with data from {\it Planck} and WMAP} \label{sec:comparison}

\subsection{Foreground model}
For the comparisons shown in this section we used the baseline foreground model described in Sec. \ref{sec:baseline}. This includes synchrotron and thermal dust with fixed spectral indices in the sky. We do not include the polarized AME component, that was not detected by {\it Planck} due to its weakness compared to the noise levels \citep{planck_2015_10}. 

\subsection{Data maps} We compared the output of our model with {\it Planck} sky observations in polarization at 30, 44, 70 and 353\,GHz, complemented by the WMAP W band at 94\,GHz and K band at 23\,GHz. In the following, we carried out the comparison between model and data smoothing to a $1^{\circ}$ common resolution, which is the resolution of our synchrotron template. Such resolution is also good for display purposes because it reduces the noise and allows an easier visual inspection of the foregrounds morphology.

The {\it Planck} frequency maps have been corrected for the polarization leakage due to bandpass mismatch with the correction maps available on the {\it Planck Legacy Archive}.  The WMAP maps have been downloaded from the LAMBDA-WMAP archive \footnote{\url{http://lambda.gsfc.nasa.gov/product/map/dr5/}}.


\subsection{Comparison with foregrounds only}
We first compared the data with a model of the sky including only the foregrounds (with no high-$\ell$ features) and without CMB and noise. In this way, we only compare the deterministic components of the model, without any random realization. The true {\it Planck} and WMAP frequency responses have been used to create the model sky as described in Sec. \ref{sec:observations}.

Figure \ref{fig:scatter_plot} shows a pixel-by-pixel comparison of true vs model $Q$ and $U$ maps. We show only the pixels inside the Galactic plane ($|b| \leq 20^{\circ}$) for $N_{\rm side}=64$ to reduce the effect of noise and CMB. Figure \ref{fig:maps_comparison} shows the maps in pseudo-colour scale, for 6 frequencies and in $Q$ and $U$ intensity. The modelled foregrounds and the observed sky have the same colour scale. As expected, the agreement is very good for the foreground-dominated frequencies.  At 70 and 94\,GHz the agreement is less good, because CMB and noise become important. 
The direct comparison of the maps shows that the foreground model is quite good once the noise is reduced (by means of degrading to $N_{\rm side}=32$). The $U$ intensity of the 94\,GHz band is noisy, which makes the comparison difficult. 

\begin{figure}
	\centering
	\includegraphics[width=1\columnwidth]{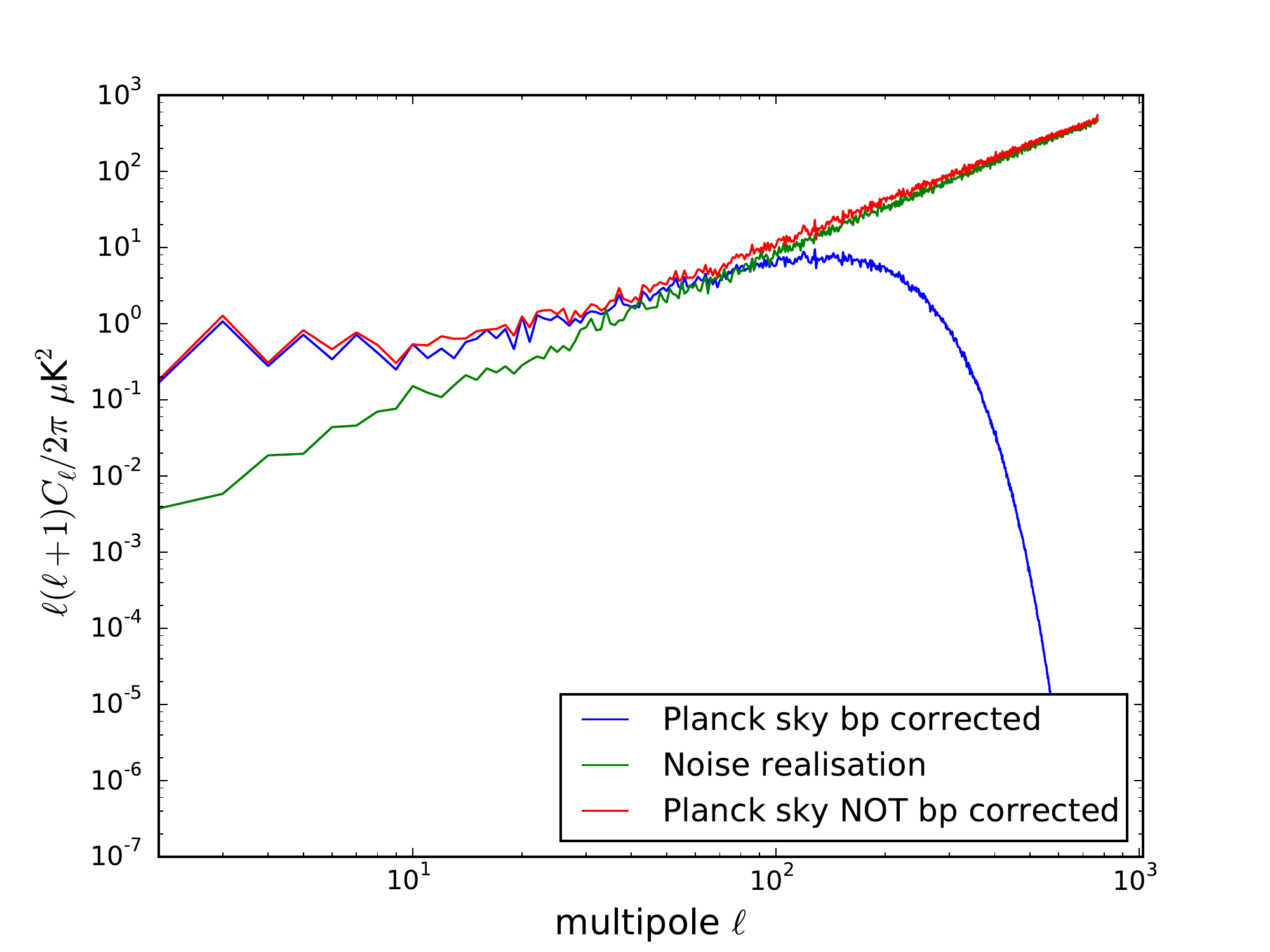}
	\caption{Full-sky $BB$ power spectra with $N_{\rm side}=256$ for both the {\it Planck} sky maps at 70\,GHz before (red) and after (blue) bandpass leakage correction. A noise realization (green) agrees well at high-$\ell$, but since the bandpass leakage correction affects the noise, there is no agreement anymore with the noise level. \label{fig:test_70}}
\end{figure}

\subsection{Including the contribution from noise and CMB} 
For the comparisons presented in this section, we included noise and a CMB realization, which are present in the sky observations, in order to assess the match of the model when all components are included. In this case we only consider the power spectra, since the different CMB and noise realizations do not allow a morphological comparison.

The CMB map has been generated starting from the best-fitting model of {\it Planck} (including polarization information, \citet{planck_2015_13}) and with tensor-to-scalar ratio $r=0.1$. 

In this case, we consider the 30, 44, 70, and 353\,GHz {\it Planck} bands and all five of the WMAP bands. The WMAP noise is simulated using the hit counts maps and RMS information available from the Lambda website. 
{\it Planck} noise has been simulated using the pixel covariance information.  However, this noise is based on the {\it Planck} data before applying the leakage correction maps, while the data used for the comparison has the correction applied. As illustrated for 70\,GHz in Fig.~\ref{fig:test_70}, the bandpass correction subtracts a large fraction of the noise, therefore the noise contribution at small scales is over-estimated. For this reason, the comparison that follows is limited to  $\ell \leq 100$ where this effect is not significant. 

Figure \ref{fig:power_spectra_complete_model_planck} shows the comparison of the $EE$ and $BB$ full-sky power spectra between the complete model (foregrounds model + noise realization + CMB realization) and the {\it Planck} bands. Figure \ref{fig:power_spectra_complete_model_wmap} shows the same for the five WMAP bands.  Since these spectra are computed over the full-sky, there is no leakage between $EE$ and $BB$ modes and no correction is needed. In this case, the error on the data power spectrum (orange shaded region) is just due to noise and CMB variance, and is calculated as
\begin{equation}
	\Delta C_{\ell} = \sqrt{\frac{2}{2\ell + 1}} (C_{\ell}^{\rm CMB}+N_{\ell}) \text{,}
\end{equation}
where $C_{\ell}^{\rm CMB}$ is the input CMB power spectrum and $N_{\ell}$ is the noise bias power spectrum, calculated from 100 noise Monte Carlo realizations based on the noise covariance matrix information of each band. This error is generally very small compared to the foreground signal ($<1$\%), and in most cases not visible in the figures. The maximum observed error, considering multipoles up to $\ell=100$, is $\sim 10$\% at 70\,GHz for \textit{Planck} and $\sim 23$\% at 61\,GHz for WMAP.

The agreement between the intermediate frequencies (44 and 70\,GHz) benefits from the inclusion of CMB and noise in the comparison. At 70\,GHz the CMB polarised intensity is strong, so the match improves. The inclusion of noise is particularly important to reconcile model and data towards $\ell = 100$.


\begin{figure}
	\centering
	\includegraphics[width=1\columnwidth]{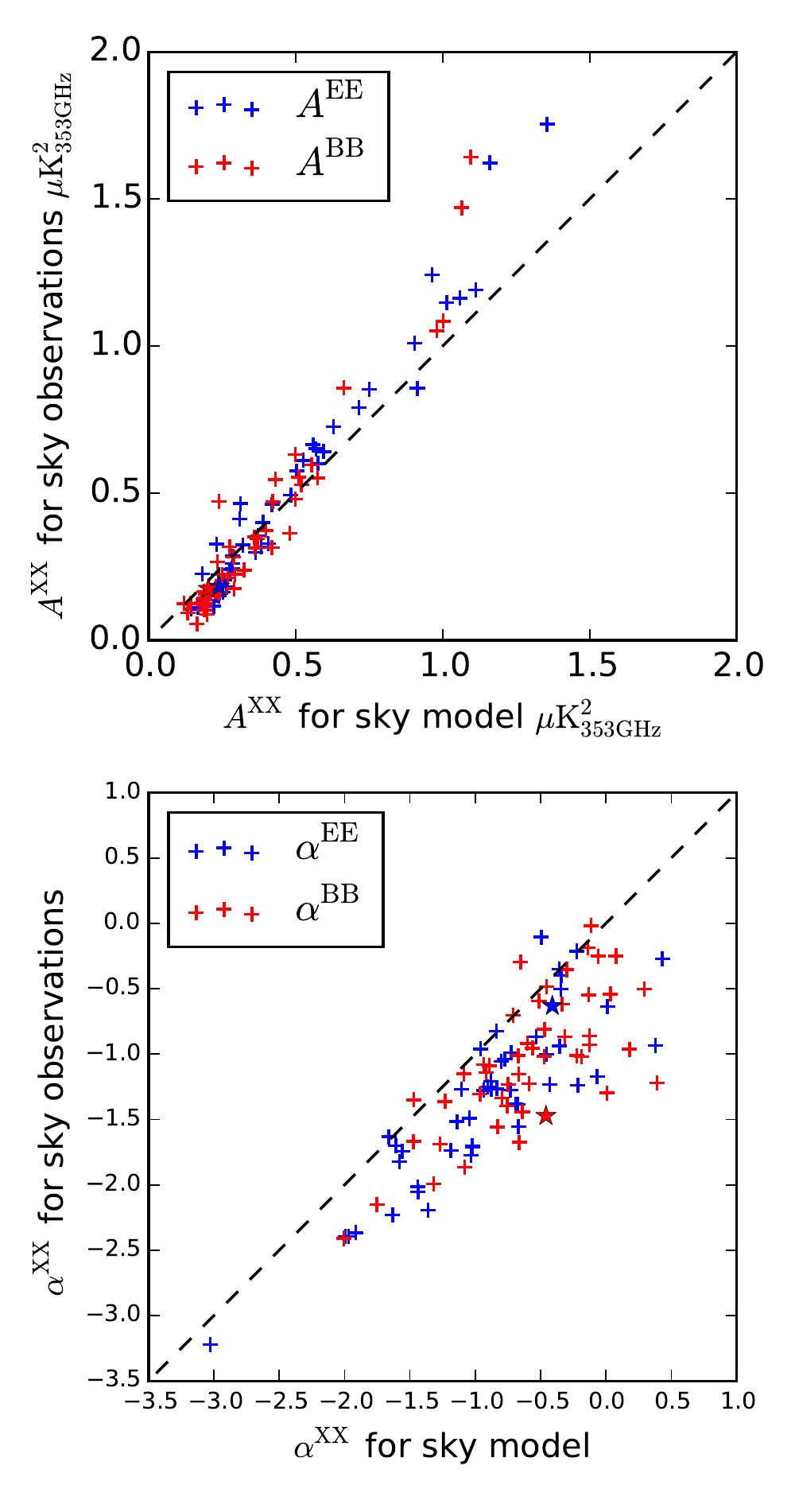}
	\caption{Correlation of the fitted power-law between our model and \textit{Planck} 353\,GHz observations for several intermediate and high latitude small patches. Top, correlation for the value of the amplitude $A^{XX}$ at $\ell=80$. The stars corresponds to the BICEP2 field. Bottom, correlation between the values of the power-law slope $\alpha_{XX}$. \label{fig:small_patches}}
\end{figure}

\subsubsection{Comparison on small patches at high Galactic latitude} As a final assessment of our model, we compare local power spectra to the latest {\it Planck} observations in several small sky patches at intermediate and high Galactic latitude. This is particularly useful for ground-based CMB polarization experiments, which target these areas. For example, the BICEP2/KECK array \citep{bicep2_paper} observes with two bands at 100 and 150\,GHz. They target a high Galactic latitude patch visible from the South Pole, with a size of $\sim 800$\,${\rm deg}^2$. The South Pole Telescope (SPT) has measured the sub-degree scales lensing $BB$ power spectrum in a southern $100$\,${\rm deg}^2$ patch using two bands (95 and 150\,GHz) \citep{south_pole_telescope}.  Another example is the POLARBEAR experiment, in the Atacama desert in Chile, which measured the lensing $BB$ spectrum in three small patches with a total area of $25 {\rm deg}^2$ at 150\,GHz \citep{polarbear}. Given the frequency coverage of these experiments, observing around 150\,GHz, where the CMB peaks, it is crucial to model correctly the contamination from polarized thermal dust emission.

For our comparison, we followed a similar procedure to the one described in \citet{planck_intermediate_30} for assessing the contamination by thermal dust. We produced disk-shaped masks with a radius of $11.3^{\circ}$ ($400$\,${\rm deg}^2$). Each patch is located on the center of a pixel of a $N_{\rm side}=8$ map and we considered patches whose center has a latitude of $|b|>45^{\circ}$. This leaves 48 circular patches. We also added another mask that selects the region targeted by BICEP2/KECK. The masks are apodized by smoothing with a $2^{\circ}$ FWHM beam.

We used the same maps assessed in  Fig.~\ref{fig:power_spectra_complete_model_planck} ($N_{\rm side}=256$, $1^{\circ}$ FWHM resolution,  modelled as  foregrounds + noise + CMB). On each small patch, we calculated the pseudo-$C_{\ell}$ for both our model and the {\it Planck} observations at 353\,GHz, in order to compare the thermal dust polarization intensity. Then, we corrected for the effect of masking with a pseudo-$C_{\ell}$ approach \citep[e.g., ][]{2005MNRAS.360.1262B}. Following \citet{planck_intermediate_30}, we fitted for the power law $D^{XX}_{\ell}=A^{XX}(\ell/80)^{\alpha_{XX}+2}$, where $XX$ is either $EE$ or $BB$, using the same fitting method described in Sec. \ref{sec:high-multipole} over a range of multipoles $\ell=40$--100.


To assess the match, we plot the fitted $A^{XX}$ and $\alpha_{XX}$ from the model and from \textit{Planck} 353\,GHz observations in Fig.~\ref{fig:small_patches}. The crosses represent each one of the disk-shaped patches, while the star represents the BICEP2 field for either $EE$ or $BB$. The agreement is better on the foreground amplitude $A^{XX}$ than on the slope $\alpha_{XX}$, our model being generally a bit steeper than the \textit{Planck}\,353 GHz channel. For the patches having lowest signal, the mismatch is mostly due to errors in modelling the noise, as detailed in Fig.~\ref{fig:test_70}. For the signal-dominated patches, the mismatch is more likely due to foreground modelling. The {\it Planck} \textit{commander} analysis that produced the templates was optimized for the full-sky rather than a small sky patch. As a consequence, the match on individual $400$\,${\rm deg}^2$ sky areas may vary, but there is a good agreement when considering a sample of sky patches.

\begin{figure}
	\centering
	\includegraphics[width=1\columnwidth]{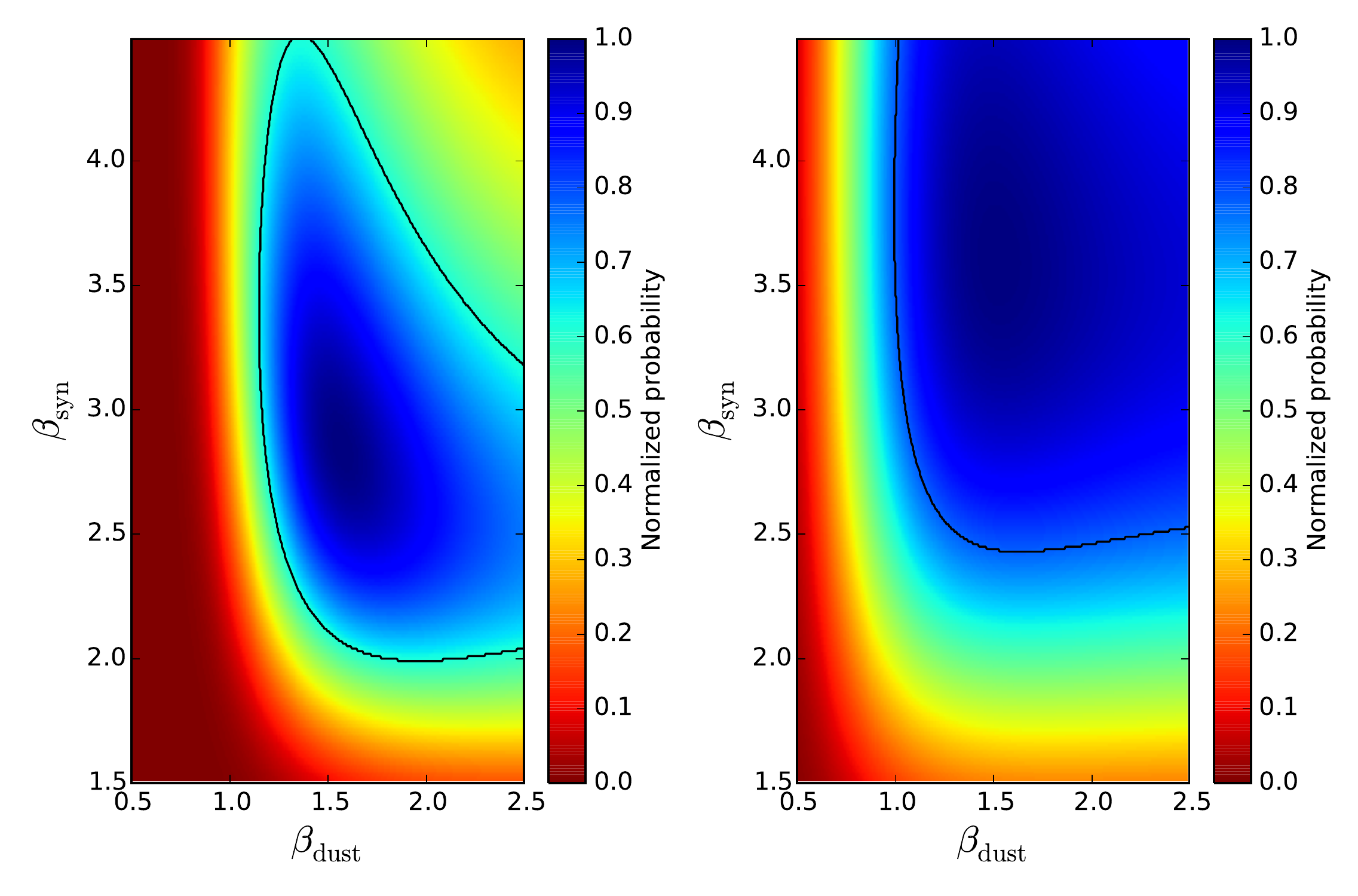}
	\caption{Normalized probability for $\beta_{\rm dust}$ and $\beta_{\rm syn}$ obtained by adding the $\chi ^2$ values for two {\it Planck} bands: 44+70\,GHz. In this case, the pixels inside a Galactic latitude $b \pm 20^{\circ}$ are used. The black curve represents the 1\,$\sigma$ confidence interval. The left panel shows the constraints given by $Q$ maps, and the right one those given by $U$ maps. \label{fig:prob_2D}}
\end{figure}

\subsection{Optimal spectral index test} In the previous tests, we used the best fit values for the spectral indices $\beta_{\rm syn}=3.10$ and $\beta_{\rm dust}=1.53$, according to \cite{planck_2015_10}. We might wonder if this is the optimal choice. Here, we explore the possibility that changing the spectral index of synchrotron and/or dust may improve the match. This analysis will also provide indications of what is a reasonable range within which to vary the synchrotron and dust spectral indices, for example for Monte Carlo purposes, while preserving a good agreement with the data. 

We do this by defining a $\chi^2$ statistic in the pixel domain. The sky observations and the model, with a common resolution of $1^{\circ}$, are degraded to $N_{\rm side}=64$. Then, we compare the maps pixel by pixel, using as the $\sigma$ error the {\it Planck} noise maps discussed in  Sec.~\ref{sec:instrumenstal_response}, properly smoothed and degraded. Therefore, for an observed band $\nu$, we minimize
\begin{equation}
	\chi^2_{\nu}(\beta_{\rm dust},\beta_{\rm syn}) = \sum_{i=0}^{N_{\rm pix}} \frac{(Q_{i,\nu}-Q_{i ,\rm model}(\beta_{\rm dust},\beta_{\rm syn}))^2}{(\sigma^{Q}_{i,\rm \nu})^2} \text{,}
\end{equation}
and analogous for $U$, where $i$ cycles through the $N_{\rm pix}$ pixels of the map being considered. 

We mapped the $\chi^2$ values for a grid of ($\beta_{\rm dust}$, $\beta_{\rm syn}$) parameters and for both the 70 and 44\,GHz bands. We could not use the {\it Planck} 30 and 353\,GHz because those are the frequencies at which the synchrotron and dust templates are normalized, therefore are unaffected by changing the spectral indices. Also, we did not consider the WMAP channels because of their higher noise levels. 
We finally obtained a joint likelihood for the 44 and 70\,GHz channels as $\exp(-\chi^2/2)$, where $\chi^2=\chi^2_{44 \rm GHz}+\chi^2_{70 \rm GHz}$. 

In Figure \ref{fig:prob_2D} we show the results for the pixels inside a $b=\pm 20^{\circ}$ Galactic strip. The left panel corresponds to the $Q$ maps, the right one to the $U$ maps. The black curve corresponds to the 1$\sigma$ confidence interval in the $\beta_{\rm dust}$-$\beta_{\rm syn}$ space considered. Notice that we assume a prior on the range of values these indices can take. The probability distribution is flatter in $U$ (and therefore the 1$\sigma$ contour is wider) because the $U$ maps have weaker foreground emission, as can be seen in Fig.~\ref{fig:maps_comparison}. 

The best fit values for the 1D marginalized probability of each parameter and its $1\,\sigma$ confidence interval, adding both bands and both polarizations ($44+70$\,GHz and $Q+U$), are $\beta_{\rm syn}=3.4\substack{+0.7\\-0.5}$ and $\beta_{\rm dust}=1.6\substack{+0.5\\-0.2}$ for the pixels inside the Galactic plane strip, and $\beta_{\rm dust}=1.6\substack{+0.6\\-0.3}$ and $\beta_{\rm syn} = 3.4 \substack{+0.9\\-0.6}$ when we consider the full-sky. The reason why we obtain relatively weak constraints is because we could only use data at central frequencies, where foregrounds are not so strong and the two spectral indices are more degenerate. A test of the model with much lower (higher) frequency would give a much stronger constraint of the synchrotron (dust) spectral index.  

\begin{figure}
	\includegraphics[width=1\columnwidth]{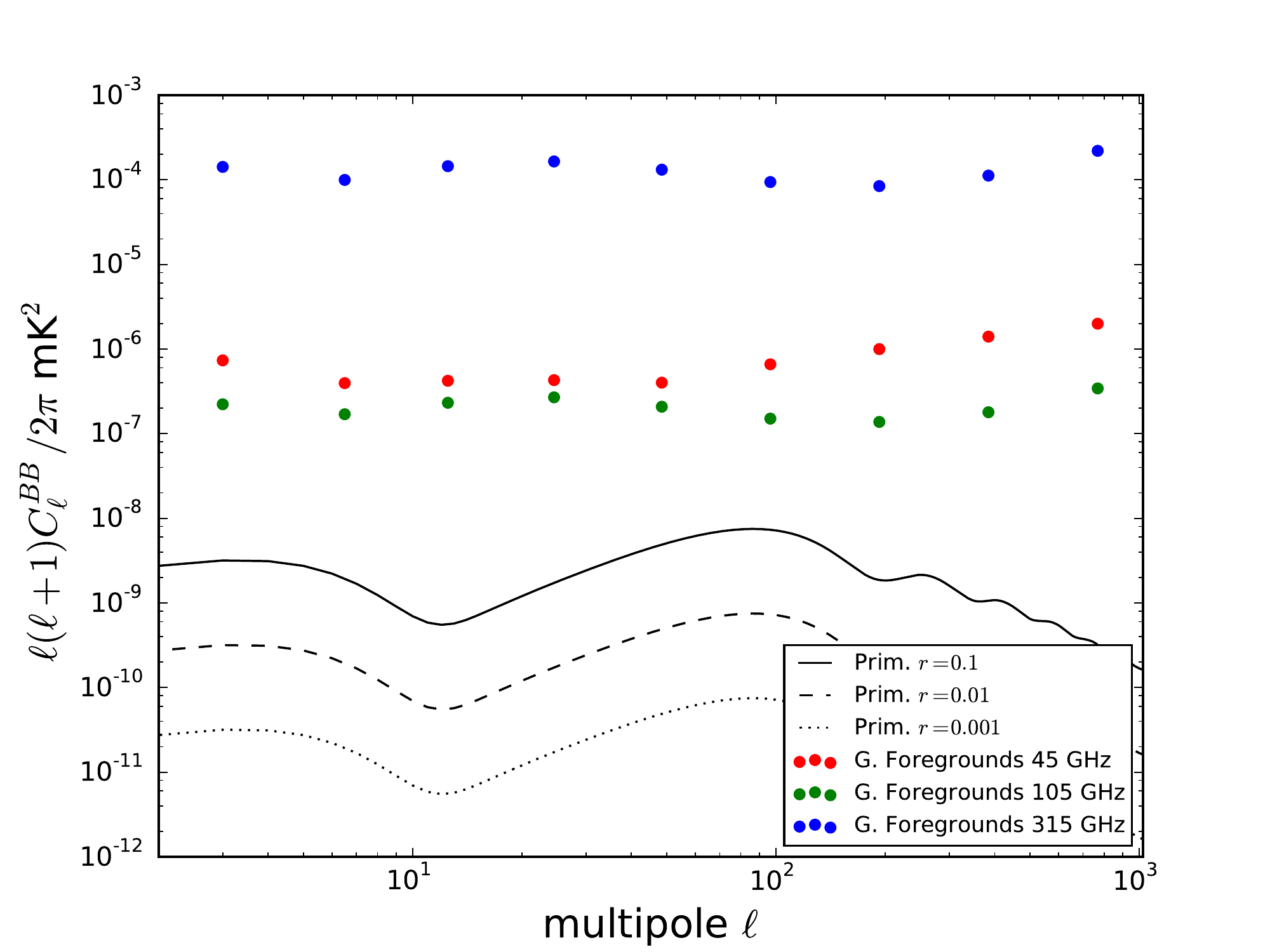}
	\caption{Forecast for primordial $BB$ modes detectability. The black curves show primordial $C_{\ell}^{BB}$ for three values of the tensor-to-scalar ratio. We also show the $BB$ power spectra for polarized foregrounds for three frequencies: 45, 105, and 315\,GHz. The foreground maps were masked with the WMAP polarization data analysis mask and deconvolved from pseudo-$C_{\ell}$ following \citet{2005MNRAS.360.1262B}. \label{fig:BB_forecast}}
\end{figure}

\begin{figure}
	\centering
	\includegraphics[width=1\columnwidth]{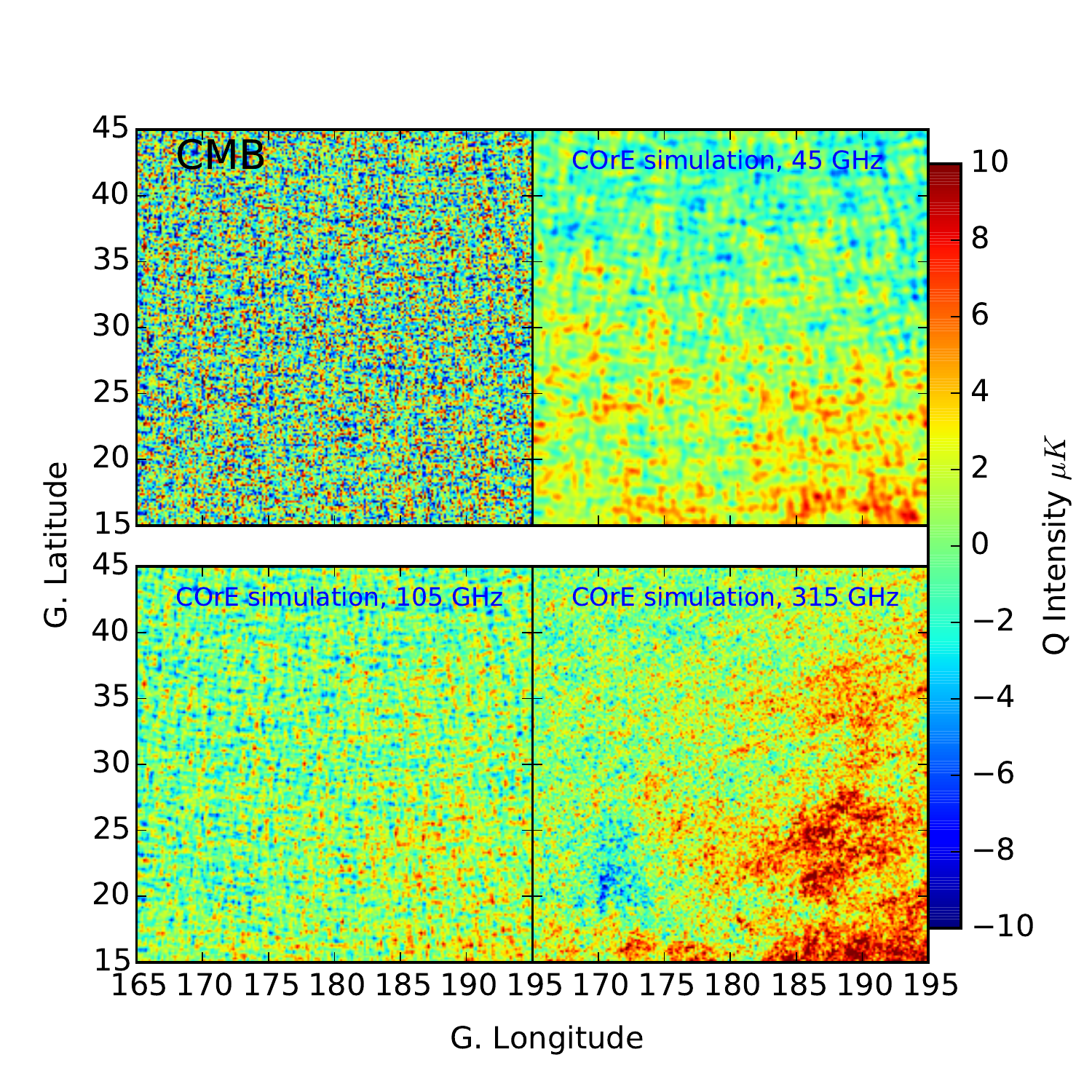}
	\caption{Zoomed-in region of intermediate foreground contaminations in a simulation of COrE observations. The maps are patches of $Q$ intensity, 30$^{\circ} \times$30$^{\circ}$, centred at $l=180^{\circ}$, $b=+30^{\circ}$. The top left panel shows a CMB realization (in thermodynamic units), and the remaining panels show the simulated observations for three frequencies: 45\,GHz (dominated by synchrotron emission), 105\,GHz (dominated by CMB), and 315\,GHz (dominated by thermal dust emission). These panels have Antenna units. \label{fig:intermediate-contamination}}
\end{figure}

\section{Forecast and Monte-carlo capabilities}\label{sec:forecast}
To illustrate the simulation capabilities of our sky model we consider the specifications of the Cosmic Origin Explorer (COrE) experiment described in \cite{core_paper}. The instrumental specifications are reported in Table \ref{table:core}. This experiment has 15 frequency bands with frequency ranging from 45 to 795\,GHz and resolution of 23--1.3\,arcminutes. The noise is simulated as Gaussian and uniform, with standard deviation of a few $\mu$K per arcminute, as quoted in the Table. 
\begin{table*}
	\centering
    \begin{tabular}{|c|c|c|c|c|c|c|c|c|c|c|c|c|c|c|c|}
	    \hline
	    Band [GHz] & 45 & 75 & 105 & 135 & 165 & 195 & 225 & 255 & 285 & 315 & 375 & 435 & 555 & 675 & 795  \\
		\hline    
	    Beam FWHM [arcmin] & 23.3&14.0&10.0&7.8&6.4&5.4&4.7&4.1& 3.7 & 3.3 & 2.8 & 2.4 & 1.9 & 1.6 & 1.3 \\
	    \hline
	    Noise [$\mu$K$\cdot$arcmin]& 8.61&4.09&3.5&2.9&2.38&1.84&1.42&2.43 & 2.94 & 5.62 & 7.01 & 7.12 & 3.39 & 3.52 & 3.60 \\
	    \hline
    \end{tabular}
    \caption{Reference values used for simulating COrE observations. \label{table:core}}
\end{table*}

Figure \ref{fig:BB_forecast} shows the forecasted $B$-mode for foregrounds for COrE as obtained with our sky model. The CMB $BB$ power spectra for different tensor-to-scalar ratios are compared with the foreground power at three COrE frequencies: 45, 105 and 315\,GHz. The foreground power spectra have been computed with the  WMAP polarization mask (excluding $\sim 37$\% of the sky); it has been corrected for again using the pseudo-$C_{\ell}$ approach of \citet{2005MNRAS.360.1262B}.  Although this figure is qualitatively similar to other forecasts in the literature \citep[e.g.][]{core_paper}, the fact that our model is based on actual polarization observations ensures a better match with the real sky and therefore improved forecast capabilities. 

In Fig. \ref{fig:intermediate-contamination} we show the $Q$ intensity on a 30$^{\circ} \times$30$^{\circ}$ sky patch located at $l=180^{\circ}$, $b=30^{\circ}$ and exhibiting intermediate foreground contamination. The top left panel shows a CMB realization with $4'$ resolution. The remaining panels show the simulated COrE observations in this region at 45\,GHz, dominated by synchrotron emission, 105\,GHz, dominated by CMB emission, and 315\,GHz, dominated by thermal dust emission. The resolution in each panel is different and corresponds to the COrE resolution quoted in Table \ref{table:core}. The features of the foreground components are easily viewed. On the smallest scales,  the foregrounds structures are those generated with the procedure described in Sec.~\ref{sec:high-multipole}, which allows overcoming the limits imposed by the intrinsic resolution of the templates (1$^\circ$ for synchrotron and $10'$ for thermal dust).  

\begin{figure}
	\centering
	\includegraphics[width=1\columnwidth]{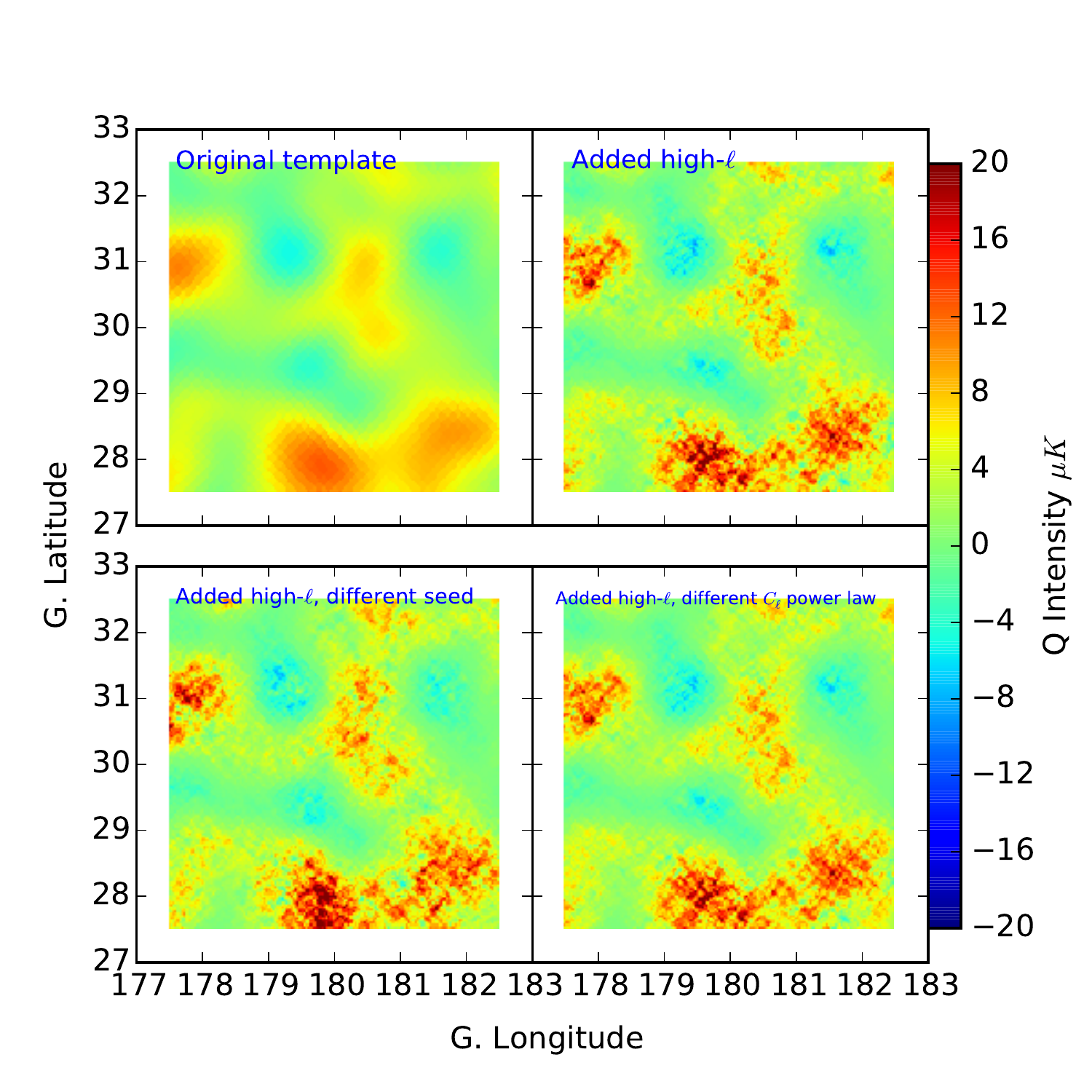}
	\caption{Illustration of the foregrounds randomization capabilities of the code on a 5$^{\circ}\times$5$^{\circ}$ sky patch centred at $l=180^{\circ}$, $b=30^{\circ}$. The top left panel shows the original $Q$ synchrotron template, which has a FWHM of $1^{\circ}$; the top right panel shows the same template upgraded in resolution to $5'$ by adding high-$\ell$ features. The bottom left panel shows the same as the top right, but with a different random realization. The bottom right panel shows the same as the top right one, but with a steeper power spectrum at high multipoles (and therefore fainter foreground features).\label{fig:montecarlo}}
\end{figure}

As mentioned, a valuable feature of our sky model is the capability to randomize to a certain degree the polarized foreground components, for example for Monte-Carlo purposes. This can be particularly useful to produce simulations that reflect our uncertainties on the polarization of the foregrounds. The randomization method exploits the same procedure described in Sec. \ref{sec:high-multipole} to upgrade the resolution of the templates. By changing the parameters describing the power spectrum  of the high-$\ell$ map (such as the slope and normalization), it is possible to obtain fainter/stronger foreground contamination at intermediate and small scales. Moreover, by changing the seed of the random realization, it is possible to obtain independent patterns for the same configuration. This is illustrated in Fig. \ref{fig:montecarlo}, where we show the $Q$ intensity for the synchrotron emission in a 5$^\circ \times$5$^{\circ}$ patch of the sky centered in the same coordinates as Fig.~\ref{fig:intermediate-contamination}. The top panels show the original synchrotron emission template at $1^\circ$ resolution on the left and a high-resolution ($5'$) one on the right. The bottom panels show two alternative versions of the top-right panel: a different realization with the same power spectrum on the left and the same random realization with steeper power spectrum (and therefore fainter foreground features) on the right.

\section{Conclusions}\label{sec:conclusions} We have constructed and validated a new model of the microwave sky in polarization, based on the most recent results from the {\it Planck} experiment \citep{planck_2015_6}. Our model features multiple choices for the frequency-dependence of the polarized synchrotron and dust foregrounds of increasing complexity. The templates, based on the {\it Planck} observations, are upgraded in resolution by means of a random map modulated by the large-scale  foreground emission. This allows simulating experiments with higher resolution than {\it Planck}. At the same time, it allows randomizing over the small-scale foreground features, for example in a Monte-Carlo approach. We also include curvature in the synchrotron spectral law, and the capability to model an arbitrary number of modified black bodies for the thermal dust law. Finally, we include the choice of a polarized AME emission template, constructed from the thermal dust polarization angles and from the AME total intensity template provided by {\it Planck}.

We demonstrated that our baseline model (power-law synchrotron with fixed spectral index $\beta_{\rm syn}=3.10$, modified blackbody thermal dust with fixed temperature $T_{\rm d}=21$\,K and spectral index $\beta_{\rm dust}=1.53$) gives a very good match with both WMAP and {\it Planck} data. We also found good agreement between the dust model and the data on small high-latitude regions, typically targeted by ground-based experiments. When changing the parameters of the model, $\beta_{\rm syn}=2.9-4.2$ and $\beta_{\rm dust}=1.4-2.1$ also provide a good fit to the data. 

We finally showed the capabilities of our model for forecast and Monte-Carlo purposes by simulating data for the COrE experiment. Our easy to use python package, which we make fully available\footnote{at \url{http://www.jb.man.ac.uk/~chervias}}, will be a useful tool for the CMB polarization community.  

\section*{Acknowledgements} CHC acknowledges the funding from Becas Chile/CONICYT. AB and MLB acknowledge support from the European Research Council under the EC FP7 grant number 280127. MLB also acknowledges support from an STFC Advanced/Halliday fellowship. We gratefully acknowledge the anonymous referee for useful suggestions that led to the improvement of this paper.



\bibliographystyle{mnras}
\bibliography{example} 








\bsp	
\label{lastpage}
\end{document}